\documentclass[journal]{IEEEtran}

\usepackage{indentfirst}
\usepackage[T1]{fontenc}
\usepackage[dvips]{graphicx}
\usepackage[dvips]{epsfig}
\usepackage{amsmath}
\usepackage{amssymb}
\usepackage{verbatim}
\usepackage{tikz} 
\usepackage{bbm}
\usepackage{algorithmic}
\usepackage{algorithm}
\usepackage{array}
\usepackage{enumerate}

\usepackage{cite}

\usepackage[linktoc=page,colorlinks=true,linkcolor=blue,citecolor=blue]{hyperref}

\newtheorem{definition}{Definition}

\newcommand{\qed}{\nobreak \ifvmode \relax \else
      \ifdim\lastskip<1.5em \hskip-\lastskip
      \hskip1.5em plus0em minus0.5em \fi \nobreak
      \vrule height0.5em width0.5em depth0.25em\fi}

\newcommand{\E}[1]{\mathbf{E}_{}\left[#1\right]}

\def\P{{\mathbf P}}

\begin{document}

\title{Simplicial Homology for Future Cellular Networks}
\author{ Ana\"is Vergne, Laurent Decreusefond, Philippe Martins,
  ~\IEEEmembership{Senior Member,~IEEE}
\thanks{Manuscript created on September 11, 2014.}
\thanks{ A. Vergne is with the Geometrica team, Inria Saclay - Ile de
  France, Palaiseau, France.} \thanks{L. Decreusefond, and P. Martins are with the Network 
and Computer Science Department, Telecom ParisTech, Paris, France.}}

\maketitle
\begin{abstract}
Simplicial homology is a tool that provides a mathematical way to
compute the connectivity and the coverage of a
cellular network without any node location information. In this
article, we use simplicial homology in order to not only compute the
topology of a cellular network, but also to discover the clusters of
nodes still with no location information. We propose three algorithms
for the management of future cellular networks.
The first one is a frequency auto-planning algorithm for
the self-configuration of future cellular networks. It aims at
minimizing the number of planned frequencies while maximizing the
usage of each one. Then, our energy conservation algorithm falls into
the self-optimization feature of future cellular networks. It
optimizes the energy consumption of the cellular network during off-peak
hours while taking into account both coverage and user
traffic. Finally, we present and discuss the performance of a disaster
recovery algorithm using determinantal point processes to patch
coverage holes.
\end{abstract}


\IEEEpeerreviewmaketitle

\section{Introduction}
\IEEEPARstart{L}{ong} Term Evolution (LTE) is the 3GPP standard specified in Releases 8 
and 9. Its main goal is to increase both capacity and speed in
cellular networks. Indeed, cellular network usage has changed over the
years and bandwidth hungry applications, as video calls, are now
common. Achieving this goal for both capacity and speed costs a lot of
money to the network operator. A solution to limit operation
expenditures is the introduction of Self-Organizing Networks
(SON) in LTE systems. 3GPP standards have indeed identified
self-organization as a necessity for future cellular networks
\cite{normelte}. Self-organization is the ability for a cellular
network to automatically configure itself and adapt its behavior
without any manual intervention. Therefore, SON features can be
divided into self-configuration, self-optimization, and self-healing
functions. We will define and describe the features we are interested
in, for a further reading a full description of SON in LTE can for
instance be found in \cite{bookson}.  

First, self-configuration functions aim at the plug-and-play paradigm: new
transmitting nodes should be automatically configured and integrated
to the existing network. Upon arrival of a new node, the
neighboring nodes update their dynamic neighbor tables thanks
to the Automatic Neighbor Relation (ANR) feature. 
Among self-configuration functions, we can find the dynamic frequency
auto-planning. It is a known problem from spectrum-sensing
cognitive radio where equipments are designed to use the best wireless
channels in order to limit interference \cite{cognitive}. The
different nodes of the secondary cognitive network have to choose the
best frequency to use in order to maximize the coverage and minimize
the interference with the base stations of the primary
network. A similar approach has been made in \cite{ofdmafemto} for the
spectrum allocation of femtocells. However, there is no strict hierarchy between the nodes of future
cellular networks, all nodes pertain to the primary network. Therefore
these solutions can not always be used here.
Moreover, while in earlier releases, static frequency planning was
preferred, it has become a critical point to allow dynamic
configuration since the network has a
dynamic behavior with arrivals and departures of base stations, and
does not always follow a regular pattern with the introduction of
femtocells and relays in heterogeneous networks. 

The second main SON feature is the category of the self-optimization
functions, which defines the ability of the network to adapt its
behavior to different traffic scenarios. Indeed, in LTE cellular
networks, eNode-Bs (eNBs) have multiple configurable parameters. An
example is output power, so cells sizes can be configured when
capacity is the limitation rather than coverage. Moreover, fast and
reliable X2 communication interfaces connect eNBs. So the whole
network has the capability to adapt to different traffic situations.
Then, users traffic can be observed via eNBs and User
Equipments (UEs) measurements. Therefore, the self-optimization functions aim
at using these traffic observations to adapt the whole network, and not
only each cell independently, to the traffic situation. 
One case where self-optimization is often needed is the adaptation to
off-peak hours. Typically a cellular network is deployed to match
daily peak hours traffic requirements. Therefore during off-peak
hours, the network is daily under-used. This leads to a huge unneeded
amount of energy consumption. An idea is thus to switch-off some of
the eNBs during off-peak hours, while other eNBs adjust their
configuration parameters to keep the entire area covered. If
the traffic grows, switched-off eNBs can be woken up to satisfy
the user demand.

The third and last of the SON main functions is self-healing. In future
cellular networks, nodes would be able to appear and disappear at any
time. Since the cellular network is not only constituted of operated
base stations anymore, the operator does not control the arrivals or
departures of nodes. But the disappearances of nodes can be more
generalized: for example in case of a natural disaster (floods,
earthquakes or tsunamis...), several nodes do disappear at once. The
self-healing functions aim at 
reducing the impacts from the failures of nodes must it be in isolated
cases, like the turning off of a Femtocells, or more serious cases
where the whole network is damaged. We are interested in this latter
case, where some of the nodes are completely
destroyed. However cellular networks are not necessarily 
built with redundancy and then can be sensitive to such
damages. Coverage holes can appear resulting in no signal  for
communication at all in a whole area. Paradoxically,
reliable and efficient communication is especially
needed in such situations. Therefore, solutions for damage recovery
for the coverage of cellular networks are much needed. 

In this article, we use simplicial homology to comply with the
self-organization requirements of future cellular networks. Simplicial
homology provides a way to represent any wireless network without any
location information, and compute its topology. A cellular network is
then represented by a combinatorial object called abstract simplicial
complex, and its topology is characterized in two dimensions by the
so-called first two Betti numbers: the number of connected components
and the number of coverage holes. But the simplicial complex
representation does not only allow the topology computation, but it
also gives geographical information, such as which nodes are in some
clusters, or which ones are more homogeneously distributed. We use
this simplicial complex representation in three algorithms that answer
three specific aspects of SON in future cellular networks. 

First, we propose a frequency auto-planning algorithm which,
for any given cellular network, provides a frequency planning
minimizing the number of frequencies needed for a given accepted
threshold of interference. The algorithm calls several instances of a
reduction algorithm, introduced in \cite{infocom}, for the
allocation of each frequency. Using simplicial complex representation
combined to the reduction algorithm allows us to obtain a 
homogeneous coverage between frequencies. 
In a second part, we enhance the reduction algorithm to satisfy any
user traffic. The reduction algorithm, as it is presented in 
\cite{infocom}, only satisfies perfect connectivity and
coverage. However, in cellular networks, especially in urban areas,
coverage is not the limiting factor, capacity is. So the optimal
solution is not optimal coverage anymore but depends on the required
traffic. We present an enhanced reduction
algorithm to reach an optimally used network. 
Finally, we present an algorithm for disaster recovery of wireless
networks first introduced in \cite{pmr}. Given a damaged cellular
network, the algorithm first adds too many nodes then runs the
reduction algorithm of \cite{infocom} to reach an optimal result. For
the addition of new
nodes we propose the use of a determinantal point process which
has the inherent ability to locate areas with low density of nodes:
namely coverage holes. 

We thoroughly evaluate the performance of 
our three homology-based greedy algorithms. We provide complexity results
and performance comparison with three graph-based greedy algorithms.
We aim at comparing our homology approach to the graph approach to see
the benefit of the use of homology. Since we propose three greedy
homology-based algorithms, then the comparison with three graph-based
algorithms is expected.  

The remainder of this article is organized as follows. After a section
on related work on self-configuration, self-optimization and recovery
in future cellular networks done in Section \ref{sec_rel}, we introduce simplicial homology
as well as the reduction algorithm we use all along the article in Section \ref{sec_prel}. Then
in Section \ref{sec_auto}, we introduce our frequency auto-planning
algorithm. The energy conservation algorithm is presented in Section
\ref{sec_opt}. We provide the disaster recovery algorithm description
in Section \ref{sec_pmr}. Finally, we conclude in Section \ref{sec_ccl}.

\section{Related work}
\label{sec_rel}
\subsection{Self-configuration in future cellular networks}
During the deployment of a cellular network, its
different nodes (eNBs, relays, femtocells) has to be configured. This
configuration happens first at the deployment, then
upon every arrival and departure of any node. The classic
manual configuration done for previous generations of cellular networks
can not be operated in future cellular networks: changes in the
network occur too often. Moreover,
the dissemination of private femtocells leads to the presence in
the network of nodes with no access for manual support. So the future
cellular networks are heterogeneous networks with no regular pattern
for their nodes. They need to be able to self-configure
themselves. The initial parameters that a node needs to configure are
its IP adress, its neighbor list and its radio access parameters. IP
adresses are out of the scope of this work, but we will discuss the
two other parameters.
The selection of the nodes to put on one's neighbor list can
be based on the geographical coordinates of the nodes and take into
account the antenna pattern and transmission power
\cite{jantti}. However, this approach does not consider changing radio
environment, and requires exact location information which can be easy
to obtain for eNBs, but not for Femtocells. The authors of \cite{kim}
propose a better criterion for the configuration of the neighbor
list: each node scans in real time the Signal to Interference plus
Noise Ratio (SINR) from other nodes, then the nodes which SINR are
higher than a given threshold are included in the neighbor list. The
neighbor list of a node is then equivalent to connectivity information
between nodes. This is the only information needed in order to build
the simplicial complex representing a given cellular network.

Among radio access parameters, we can find frequency but also
propagation parameters since the apparition of beamforming techniques
via MIMO. Let us focus on the former which is the subject of Section
\ref{sec_auto}.  The frequency planning problem was first
introduced for GSM networks. However the constraints were not the same: the
frequency planning was static with periodic manual optimizations, and
in simulations, base stations were regularly deployed along an hexagonal 
pattern. With the deployments of Femtocells, outdoor relays, and
Picocells, future cellular networks vary from GSM network in two major
points. First, cells do not follow a regular pattern anymore, then
they can appear and disappear at any time. Therefore the 
frequency planning problem has to be rethought in an automatic way.  A
naive idea for frequency auto-planning would simply be applying the
greedy coloring algorithm to the sparse interference graph \cite{irani}.
However, even if the provided solution may be optimal for the number of
needed frequencies, the utilization of each frequency can be disparate: one
can be planned for a large number of nodes compared to another
planned for only few of them. Then if the level of interference
increase (more users, or more powered antennas), this could lead to
communication problems for the over-used frequency, and a whole new
planning is needed. On the contrary, a more homogeneous resource
utilization can be more robust if interference increase, since there
are less nodes using the same frequency on average.
We provide here a frequency auto-planning algorithm which aims at a
more homogeneous utilization of the resources.
Moreover, the planning of frequency channels for new nodes that do
not interfere with existing nodes while still providing enough bandwidth
is still an open problem. It has been addressed in the cognitive radio
field, but these algorithms usually enable opportunistic spectrum
access \cite{brian}. However, it is not possible to extend this type
of algorithm to the frequency allocation of new nodes in cellular
networks. Indeed, the new nodes would be part of the primary network,
with a quality of service to achieve, so their frequency allocation
needs to be guaranteed and not opportunistic. In \cite{ofdmafemto},
the authors propose a spectrum allocation for femtocells in a cellular
network that is more suited to our needs. However, the frequency
planning of the femtocells occurs after the frequency planning of the
main cells (eNBs). We propose an algorithm that do not distinguish
between different types of cells. Indeed there are more types
of cells than exactly two, relays fall in between and femtocells are
not necessarily alike. In our algorithm, the planning of all the
nodes: eNBs, relays or femtocells, is done together.

\subsection{Self-optimization in future cellular networks}
In order to ensure that future cellular networks are still efficient
in terms of both Quality of Service (QoS) and costs, the self-configuration
is not sufficient. Indeed, future cellular networks have the ability
to adjust their parameters to match different traffic situation. 
Periodic optimization based on log reports, and operated centrally is
not an effective solution in terms of speed and costs. That is why we
need self-optimization. Self-optimization can be classified in three
types depending on its goal. 
First we can consider load balancing optimization. There is
multiple ways to adapt a cellular network to different loads: it is
for example possible to adapt the resources available in different
nodes. These schemes were mainly introduced for GSM \cite{das},
and then CDMA \cite{agha}, but the universal frequency reuse of LTE
and LTE-Advanced diminishes their applicability. Then one can adapt
the traffic strategy with admission controls on given cells and forced handovers
\cite{fujii}. However, as the previous solution, it is not very
suitable for OFDMA networks which require hard handover. Finally it is
possible to modify the coverage of a node by changing either its antennas
radiation pattern \cite{du} or the output power
\cite{rittenhouse}. We use this latter approach to reach an optimal
result: we adapt the coverage radius of each node
to be the minimum required to cover a given area.

The second type self-optimization is the capacity and coverage
adaptation via the use of relay nodes \cite{peng}, while the third is
interference optimization. Our energy conservation algorithm presented
in Section \ref{sec_opt} could lie in this third
category as the simplest approach towards interference control is
switching off idle nodes.  It is done based on cell
traffic for Femtocells in \cite{claussen}: after a given period of
time in idle mode, the node puts itself on stand-by. However, if one
wants to take into account the whole network, it has to consider the
coverage of the network before disconnecting, which is not the case of
Femtocells, which are by definition redundant to the base stations
network. Without considerations of traffic, we proposed in
\cite{infocom} an algorithm that reduces power consumption in wireless 
networks by putting on stand-by some of the nodes without impacting the
coverage. We can also cite \cite{nash} that proposes a game-theoretic
approach in which nodes are put on stand-by according to a coverage
function, but unmodified coverage is not guaranteed. In both these
works, only coverage is taken into account. This approach could 
eventually fit the requirements of cellular network in non-urban
cells, if their deployment has coverage redundancy. But it is not
valid for urban cells, where it is not coverage but capacity that
delimits cells. Our present algorithm goes a little
bit further by adapting the switching-off of the nodes to the whole
network situation, considering both traffic and coverage.

\subsection{Recovery in future cellular networks}
The first step of recovery in cellular networks is the
detection of failures. The detection of the failure of a cell occurs when its
performance is considerably and abnormally reduced. In \cite{kaschub},
the authors distinguish three stages of cell outage: degraded,
crippled and catatonic. This last stage matches with the event of a
disaster when there is complete outage of the damaged cells.
After detection, compensation from other nodes can occur through relay assisted
handover for ongoing calls, adjustments of neighboring cell sizes via
power compensation or antenna tilt. In \cite{amirijoo}, the authors
not only propose a cell outage management description but also
describe compensation schemes. These steps of monitoring and
detection, then compensation of nodes failures are comprised under the
self-healing functions of future cellular networks described in
\cite{selfhealing}.  

In Section \ref{sec_pmr}, we are interested in what happens when self-healing is
not sufficient. In case of serious disasters, the compensation from
remaining nodes and traffic rerouting might not be sufficient to provide service
everywhere. In this case, the cellular network needs a manual
intervention: the adding of new nodes to compensate the failures of
former nodes. However a traditional restoration with brick-and-mortar
base stations could take a long time, when efficient communication is
particularly needed. In these cases, a recovery trailer fleet of base
stations can be deployed by operators \cite{att}, it has been for
example used by AT\&T after 9/11 events. But a question remains:
where to place the trailers carrying the recovery base stations. An
ideal location would be adjacent to the failed node. However, these
locations are not always available because of the disaster, plus the
recovery base stations may not have the same coverage radii than the
former ones. Therefore a new deployment for the recovery base stations has
to be decided, in which one of the main goal is complete coverage of
damaged area. This can be viewed as a mathematical set cover problem,
where we define the universe as the area to be covered
and the subsets as the balls of radii the coverage radii. Then the
question is to find the optimal set of subsets that cover the
universe, considering there are already balls centered on the existing 
nodes. It can be solved by a greedy algorithm \cite{greedy}, $\epsilon$-nets
\cite{epsilon-nets}, or furthest point sampling \cite{fps2,fps1}. But
these mathematical solutions provide an optimal mathematical result
that do not consider any flexibility at all in the choosing of the new
nodes positions, and that can be really sensitive to imprecisions in
the nodes positions.

For a further reading, a complete survey on SON for future cellular
networks is given in \cite{surveyson}. 

\section{Preliminaries}
\label{sec_prel}
\subsection{Simplicial homology}
First we need to remind some definitions from simplicial homology for a
better understanding of the simplicial complex representation of
cellular networks.

When representing a cellular network with only connectivity
information (i.e. neighbors lists) available, one's first idea will be
a neighbor graph, where nodes are represented by vertices, and an
edge is drawn whenever two nodes are on each other neighbors
list. However, the graph representation has some limitations; first
of all there is no notion of coverage. Graphs can be generalized to
more generic combinatorial objects known as simplicial
complexes. While graphs model binary relations, simplicial complexes
represent higher order relations. A simplicial complex is a
combinatorial object made up of vertices, edges, triangles,
tetrahedra, and their $n$-dimensional counterparts. Given a set of
vertices $V$ and an integer $k$, a $k$-simplex is an unordered subset
of $k+1$ vertices $[v_0,v_1\dots, v_k]$ where $v_i\in V$ and
$v_i\not=v_j$ for all $i\not=j$. Thus, a $0$-simplex is a vertex, a 
$1$-simplex an edge, a $2$-simplex a triangle, a $3$-simplex a
tetrahedron, etc. 

Any subset of vertices included in the set of the $k+1$ vertices of a
$k$-simplex is a face of this $k$-simplex. Thus, a $k$-simplex has
exactly $k+1$ $(k-1)$-faces, which are $(k-1)$-simplices. For example,
a tetrahedron has four $3$-faces which are triangles. A simplicial
complex is a collection of simplices which is closed with respect to
the inclusion of faces, i.e. all faces of a simplex are in the set of
simplices, and whenever two simplices intersect, they do so on a
common face. An abstract simplicial complex is a purely combinatorial
description of the geometric simplicial complex and therefore does not
need the property of intersection of faces. For details about
algebraic topology, we refer to \cite{hatcher}.

Given an abstract simplicial complex, its topology can be 
computed via linear algebra computations. The so-called Betti numbers
are defined to be the dimensions of the homology groups and are easily
obtained by the rank-nullity theorem, of which a proof is given in
\cite{hatcher}. But the Betti numbers also have a geometrical
meaning. Indeed, the $k$-th Betti number of an abstract simplicial
complex $X$ is the number of $k$-th dimensional holes in $X$. 
In two dimensions we are only interested in the first two Betti numbers: $ \beta_0$
counts the number of $0$-dimensional holes, that is the number of
connected components, and $\beta_1$ counts the number of holes in the
plane, i.e. coverage holes. Therefore computing the Betti numbers of
an abstract simplicial complex representing a cellular network gives
the topology of the initial network. 
For the remainder of the paper we may drop the adjective ``abstract''
from abstract simplicial complex, since every simplicial complex in
this paper is abstract.

\subsection{Reduction algorithm}
In this section, we recall the steps of the reduction algorithm
for abstract simplicial complexes presented in \cite{infocom} that we
will use all along this article. The algorithm takes
as input an abstract simplicial complex: here it is the complex
representing the cellular network, and a list of boundary vertices
that can be given by the network operator. This list of boundary
vertices is needed in order to define the area of the network and not
shrink it during the process of reduction.
Then the goal of the reduction algorithm is to remove vertices form
the abstract simplicial complex without modifying its Betti
numbers. That translates to a network by turning off nodes from the
network without modifying nor its connectivity neither its coverage.

To cover an area, only $2$-simplices are needed. So the first step of
the reduction algorithm is to characterize the superfluous
$2$-simplices of the complex for its coverage.  To do that, we define
a degree for every $2$-simplex: 
\begin{definition}
  We define the degree of a $2$-simplex $[v_0,v_1,v_2]$
  to be the size of the largest simplex it is part of:
  \begin{eqnarray*}
    D[v_0,v_1,v_2]=\max\{ d \mid [v_0,v_1,v_2] \subset d\text{-simplex}\}.
  \end{eqnarray*}
\end{definition}
For future algorithms descriptions, we will simply denote
$D_1(X),\dots,D_{s_{2}}(X)$ the $s_2$ degrees of the $s_2$
$2$-simplices of the complex $X$, with $s_k$ being the number of
$k$-simplices of $X$.

Next, in order to remove
vertices, and not $2$-simplices, we need to transmit the
superfluousness information of its $2$-cofaces ($2$-simplices it is a
face of) to a vertex via what is called an
index. An index of a vertex is defined to be the minimum of the
degrees of the $2$-simplices it is a face of. Indeed, a vertex is as
sensitive for the coverage as its most sensitive $2$-simplex. The
boundary vertices are given a negative index to mark them as 
unremovable by the algorithm: we do not want the covered area to be
shrunk.
\begin{definition}
  The index of a vertex $v$ is the minimum of the degrees of the
  $2$-simplices it is a vertex of:
  \begin{eqnarray*}
    I[v]=\min\{ D[v_0,v_1v_{2}] \mid v \in [v_0,v_1,v_2]\},
  \end{eqnarray*}
If $v$, a vertex, is a boundary vertex, then $I[v]=-1$.
\end{definition}

Finally, the indices give an optimal order for the removal of the vertices: the
greater the index of a vertex, the bigger the cluster it is part of,
and the more likely it is 
superfluous for the coverage of its abstract simplicial complex. 
 Therefore, the vertices with the greatest index are
candidates for removal: one is chosen randomly. If its removal does
not change the homology, i.e. if it does not modify its Betti numbers
$\beta_0$ and $\beta_1$, then it is effectively removed. Otherwise it
is flagged as unremovable the same way the boundary vertices are. The
algorithm goes on until every remaining vertex 
is unremovable, thus achieving optimal result. 

For more information on
the reduction algorithm we refer to \cite{infocom}.

\subsection{Simulation model}
We want to represent the nodes of a cellular network and its
coverage. We are interested in future cellular networks equipped with
SON technology, that means cellular networks of $4$-th and higher
generation. For example for a LTE network, its nodes are the eNBs,
femtocells, and relays that constitute it. 
Then we want to compute the network's coverage
constituted of coverage disks centered on the nodes. 
The \u{C}ech abstract simplicial complex provides the exact
representation of the network's coverage. Its construction for a
fixed coverage radius $r$ for all the network's nodes is given:
\begin{definition}[\u{C}ech complex]
  Given $(X,d)$ a metric space, $\omega$ a finite set of points in
  $X$, and $r$ a real positive number. The
  \u{C}ech complex of  $\omega$, denoted
  $\mathcal{C}_r(\omega)$, is the abstract simplicial complex
  whose $k$-simplices correspond to $(k+1)$-tuples of vertices in
  $\omega$ for which the intersection of the $k+1$ balls of radii
  $r$ centered at the $k+1$ vertices is non-empty.
\end{definition}

However, the \u{C}ech complex can be hard to compute, and requires
some geographical information that is not always available. For
instance Femtocells are not GPS-enabled. There exists an approximation
of the coverage \u{C}ech complex that is only based on the
connectivity information: the so-called neighbor list of each nodes of
a SON-capable cellular network. This approximation is the Vietoris-Rips abstract simplicial
complex which is defined as follows:
\begin{definition}[Vietoris-Rips complex]
  Given $(X,d)$ a metric space, $\omega$ a finite set of points in
  $X$, and $r$ a real positive number.  The Vietoris-Rips
  complex of parameter $2r$ of $\omega$, denoted
  $\mathcal{R}_{2r}(\omega)$, is the abstract simplicial complex
  whose $k$-simplices correspond to unordered $(k+1)$-tuples of
  vertices in $\omega$ which are pairwise within distance less than
  $2r$ of each other.
\end{definition}

The definitions of the \u{C}ech and the Vietoris-Rips
complex of a cellular network can be extended to include different
distance parameters in order to represent nodes with different
coverage radii. 
The \u{C}ech complex represents the network and its exact
topology. However,  using the Vietoris-Rips
complex representation, it is possible to have so-called triangular
holes in the network that do not appear in the complex. The
probability of that happening in computed in \cite{Feng2}, and in our
case where it is upper-bounded by about $0.03\%$ for a cellular network
simulated with a Poisson point process.

In simulations and figures, we can consider either a
communication/coverage disk approach, or directly a neighbor list approach.
In the first case, any node within the
communication disk of a given node is added to its neighbor
list, or when two coverage disks of two nodes intersect they are added to each
other neighbor lists. Note that communication radii are
larger, usually twice as large, than coverage radii. Finally, the abstract simplicial complex of Vietoris-Rips type is
then build based on the neighbor lists.

\section{Self-configuration frequency auto-planning algorithm}
\label{sec_auto}
\subsection{Problem formulation}
In the frequency planning problem, the topology of the network is not
relevant since it is not modified. So we use the simplicial complex
representation only for the characterization of clusters without
location information, and not for topology computation.
First we need to build the abstract simplicial complex representing
the cellular network.
In future cellular networks, every transmitting nodes (eNBs, Femtocells,
relays...) have a neighbor list created and updated with the ANR
feature. 
 With this neighbor list information, we can build the abstract simplicial complex
representing the network, each node is represented by a $0$-simplex
and each neighbor, either $1$-way or $2$-way, relationship is
represented by a $1$-simplex. The other simplices are then created
with only the $1$-simplices information (when three $0$-simplices are
connected via three $1$-simplices then a $2$-simplex is created, and
so on).

The goal of a frequency planning algorithm is to assign frequencies to
every network's nodes so that the interference between
them is minimum using the smallest number of frequencies possible. 
In this article, we only consider the one frequency per node case, and
co-channel interference, i.e.  interference between two nodes using the same
frequency. However, the main idea of the algorithm can be extended to
several frequencies per node, and interference between different
frequencies by considering group of frequencies.

Interference is a two nodes relationship so it can be represented by an
interference graph. It is possible to consider any interference
model steady through frequencies and time (at least the duration of
the configuration). Reliable communication is achievable if the interference
is under a chosen threshold. In the interference graph, every network
node is represented, then if the interference between two nodes is
higher than the threshold, an edge is drawn between
them. Consequently, two nodes linked by an edge in the interference 
graph shall not share the same frequency or the interference level will
be too high for reliable communication inside at least one of the two
cells. 

\subsection{Algorithm description}
\begin{figure}[h]
      \scalebox{0.29}{\includegraphics{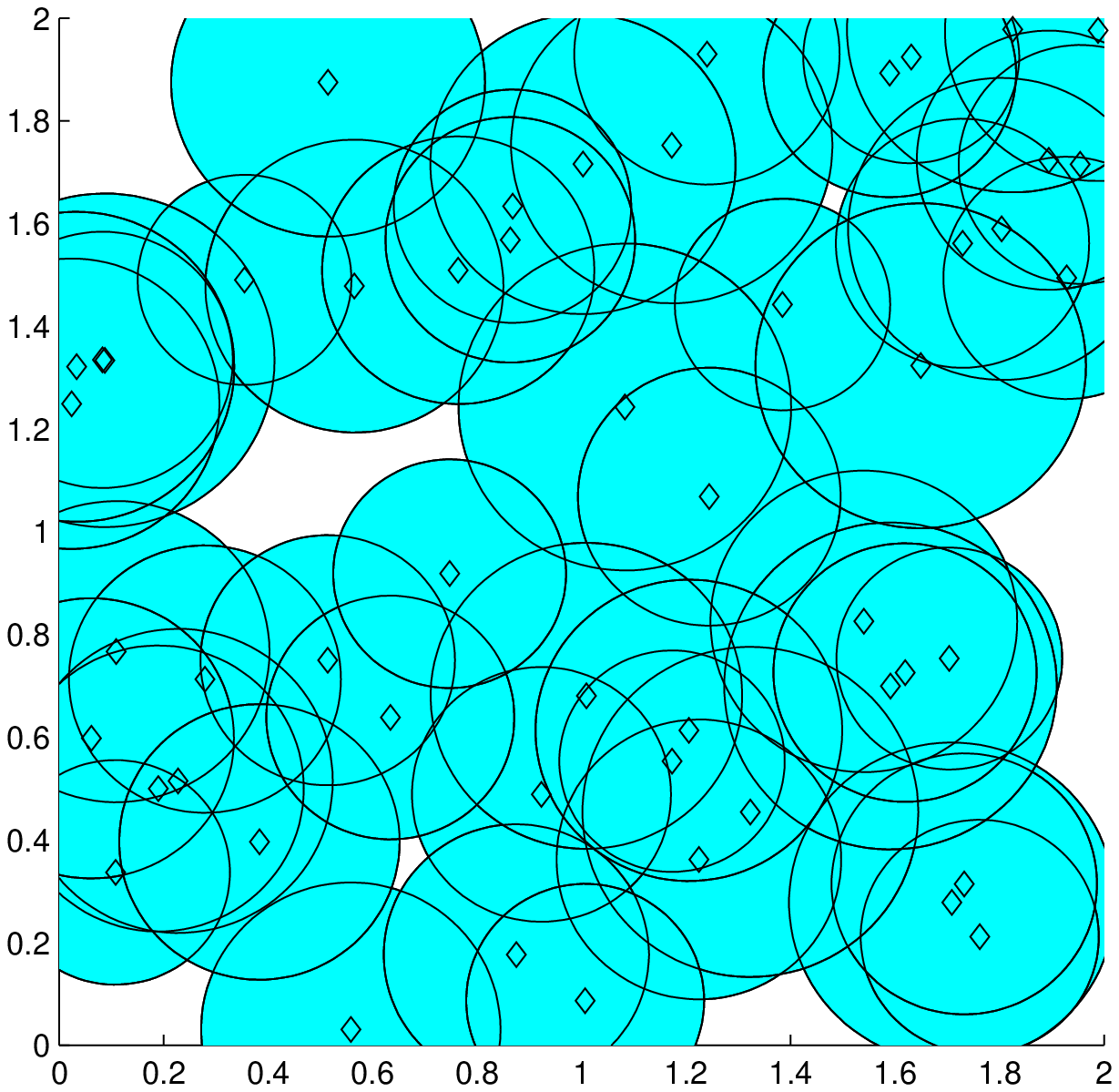}} 
\hfill
      \scalebox{0.29}{\includegraphics{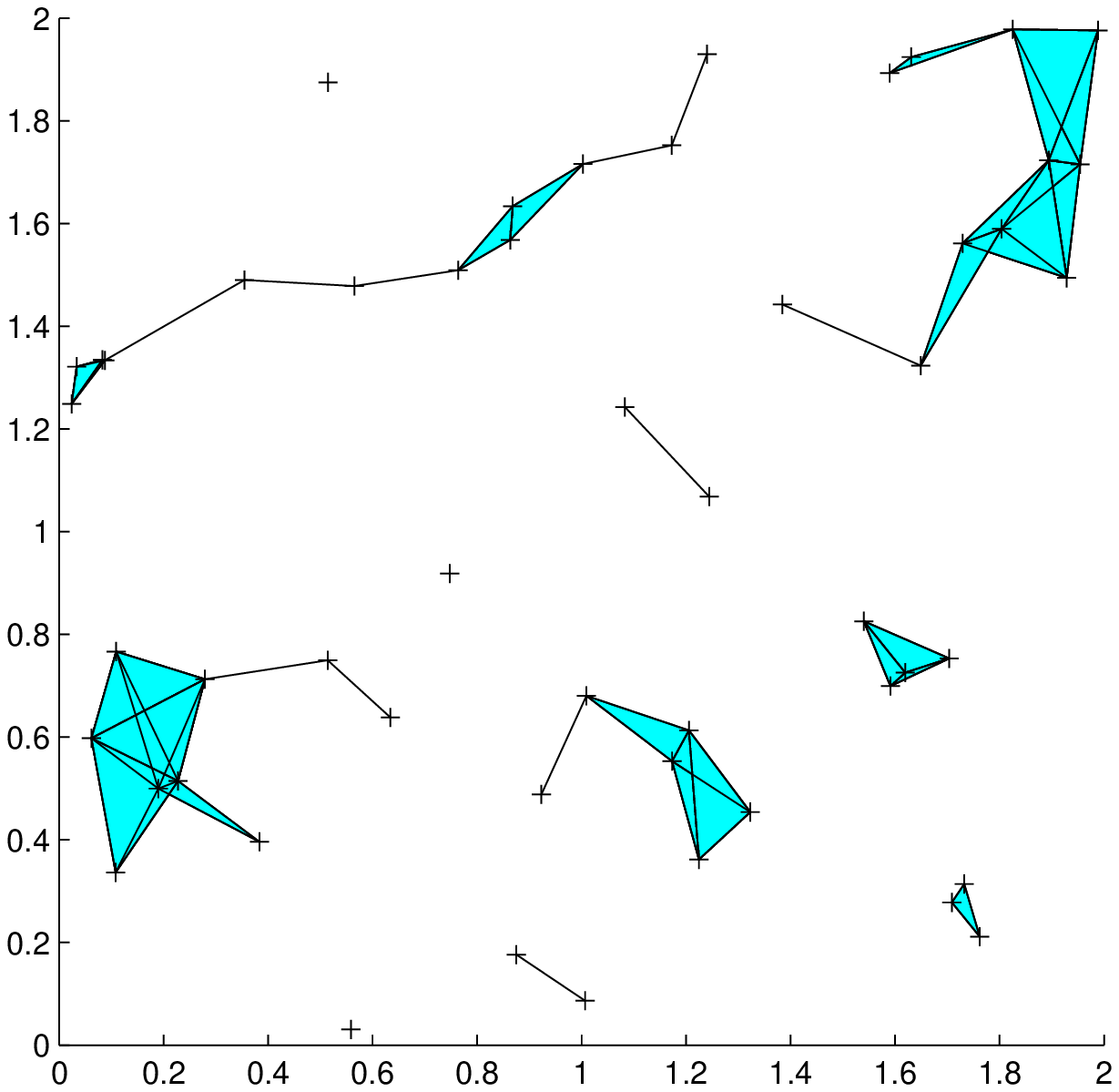}}
\caption{A cellular network and its abstract simplicial complex representation.}
    \label{fig_celco}
  \end{figure}

\begin{figure}[h]
\scalebox{0.29}{\includegraphics{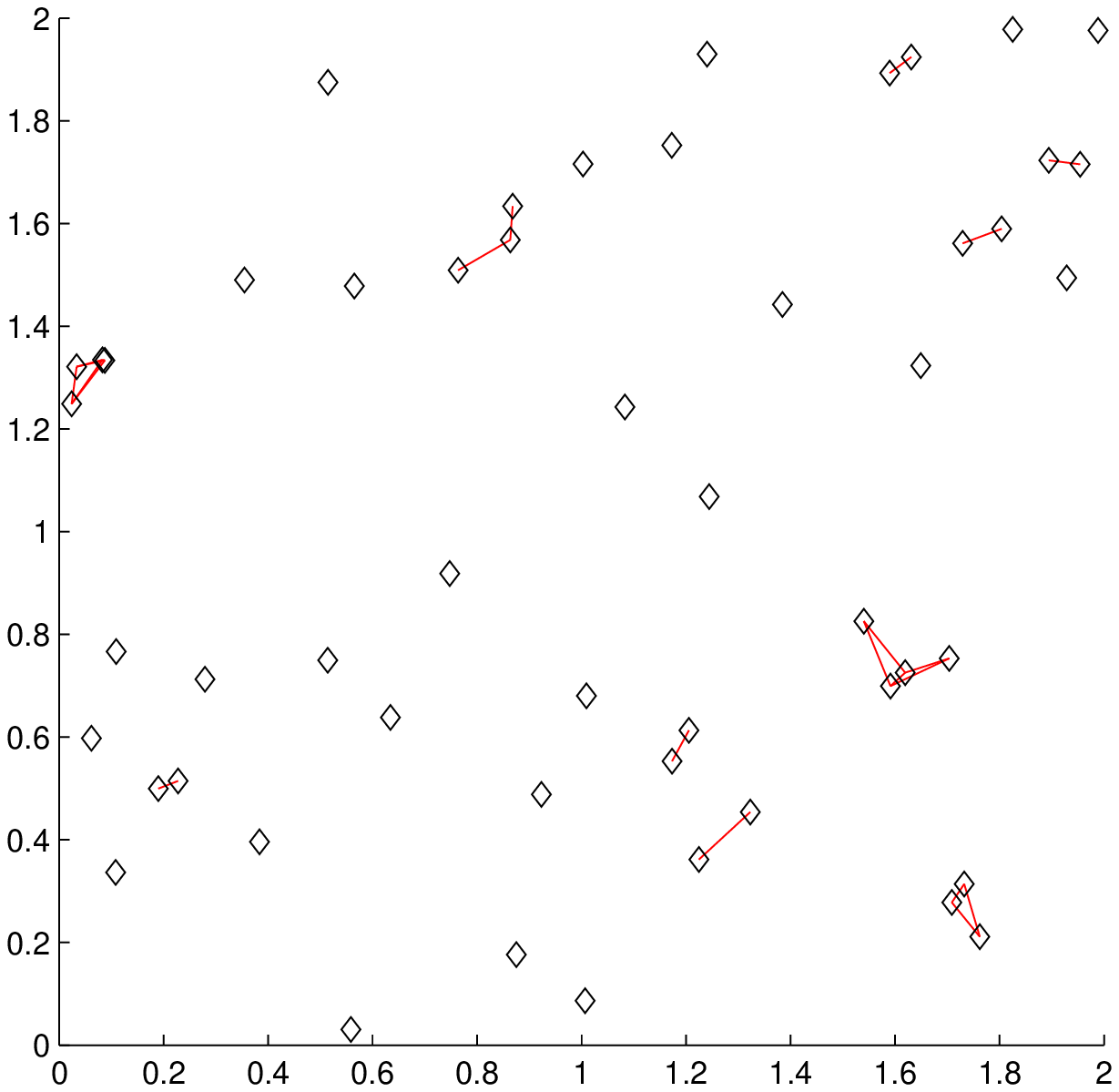}}
\hfill
      \scalebox{0.29}{\includegraphics{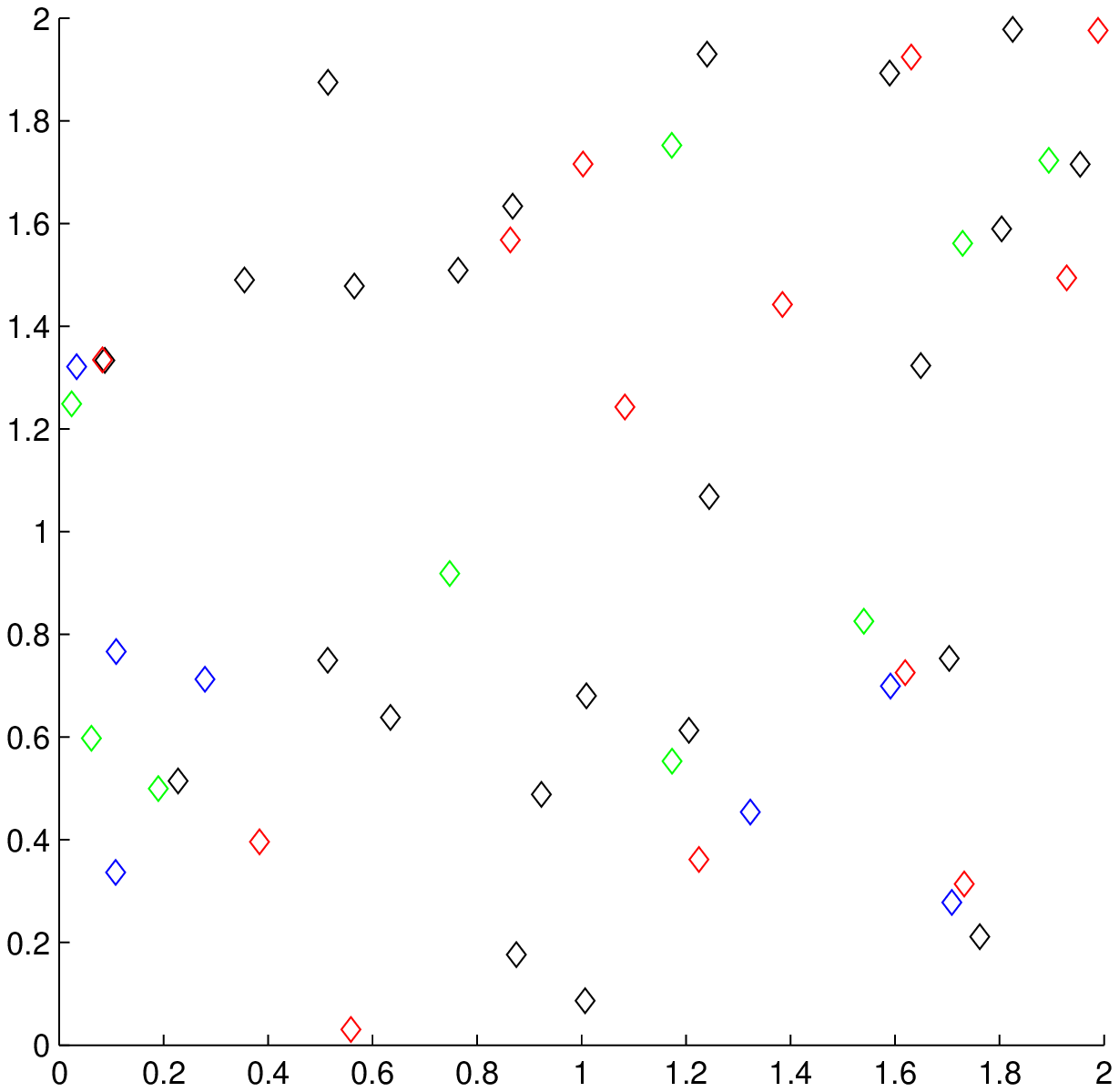}}
\caption{Interference graph and frequency planning scheme.}
    \label{fig_planning}
  \end{figure}

We consider a cellular network, nodes and communication radii, as we
can see an example in Fig. \ref{fig_celco}, of which
we compute its abstract simplicial complex
representation based on the neighbor lists. 
In Fig. \ref{fig_planning}, we can see the interference graph on the left. In our
example we choose the interference to be only distance-based. In the
general case, the only requirement for the interference model is that
it can be represented by a graph constant throughout the frequencies
and the duration of the configuration. 
We can see that the configuration of interference of Fig.
\ref{fig_planning} will need $4$ frequencies because of the $4$ linked nodes
on the left.

The goal of the algorithm is than to assign a frequency to each node
so that no to nodes with interference share the same frequency. The
algorithm begins by selecting the nodes that will receive the first
frequency available. To do that, we apply
a modified version of  the reduction algorithm presented in Section
\ref{sec_prel}. It ``removes'' nodes until obtaining a
interference-free configuration of nodes that receive the first frequency.
The order in which the nodes are removed is still decided by the indices but the
stopping condition is not the same as in simple reduction. Instead
of stopping when the maximum index among every remaining nodes is
below a given number, the algorithm stops when there is no
more pair of nodes connected to one another in the interference
graph. The resulting nodes of the reduction algorithm are then
assigned the first frequency and put aside for the remaining of the
algorithm.

All the previously ``removed'' nodes are then collected, and the corresponding
abstract simplicial complex recovered. This complex is a subset of
the initial complex so there is no need to build another
one from scratch. The next step is then to reapply the
modified reduction algorithm to this recovered 
complex to obtain a second set of nodes to which we assign the
second frequency. The algorithm goes on until every node has an
assigned frequency. 

At the end, we have a frequency assigned to every node. We ensured
that no two nodes sharing the same frequency will be too close to
each other: interference will be under a given threshold. Moreover,
the use of our coverage reduction 
algorithm with the optimized order for nodes removal allows us to
obtain a homogeneous usage of every frequency.

\begin{figure}[h]
      \scalebox{0.29}{\includegraphics{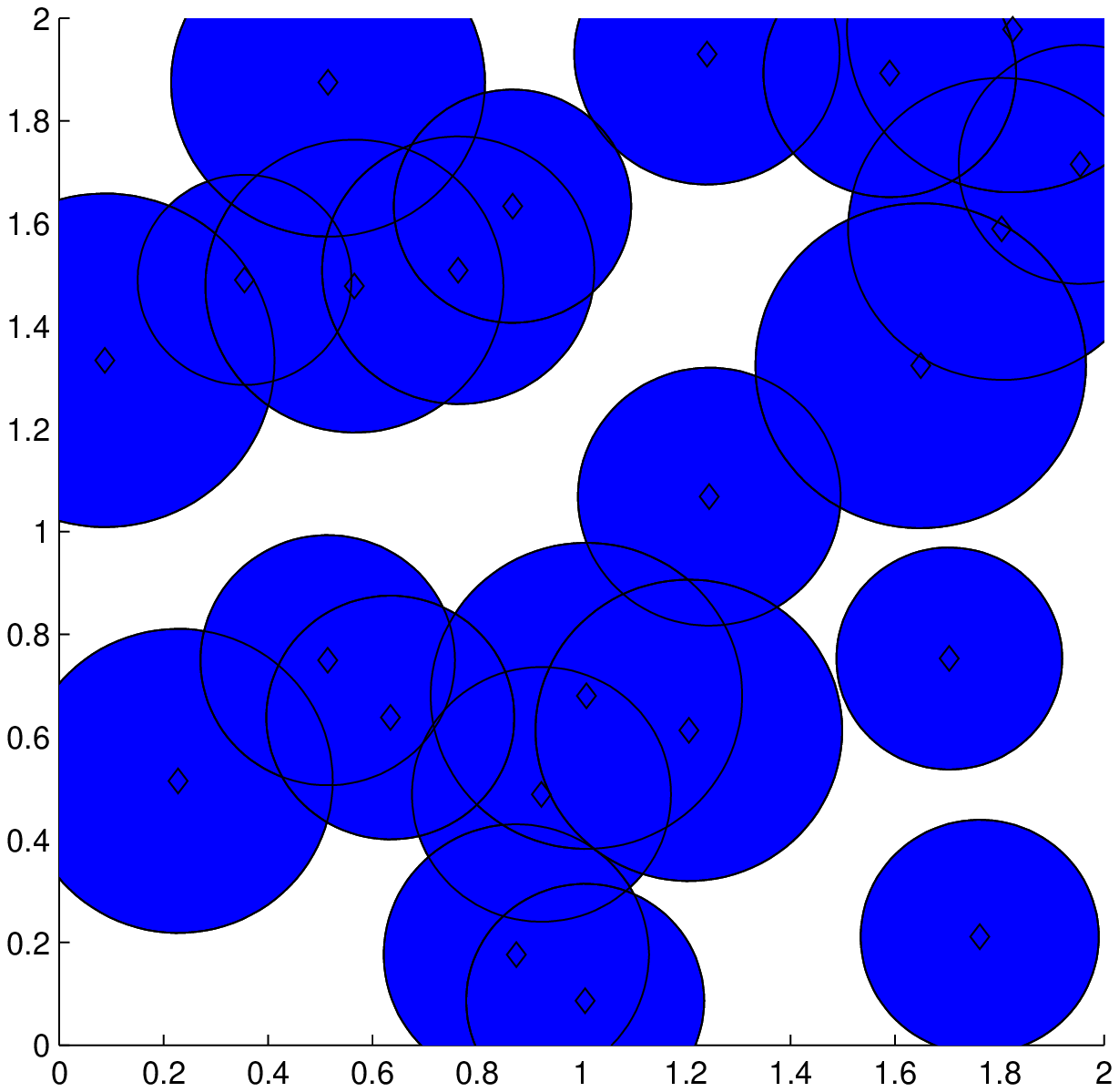}} 
      \scalebox{0.29}{\includegraphics{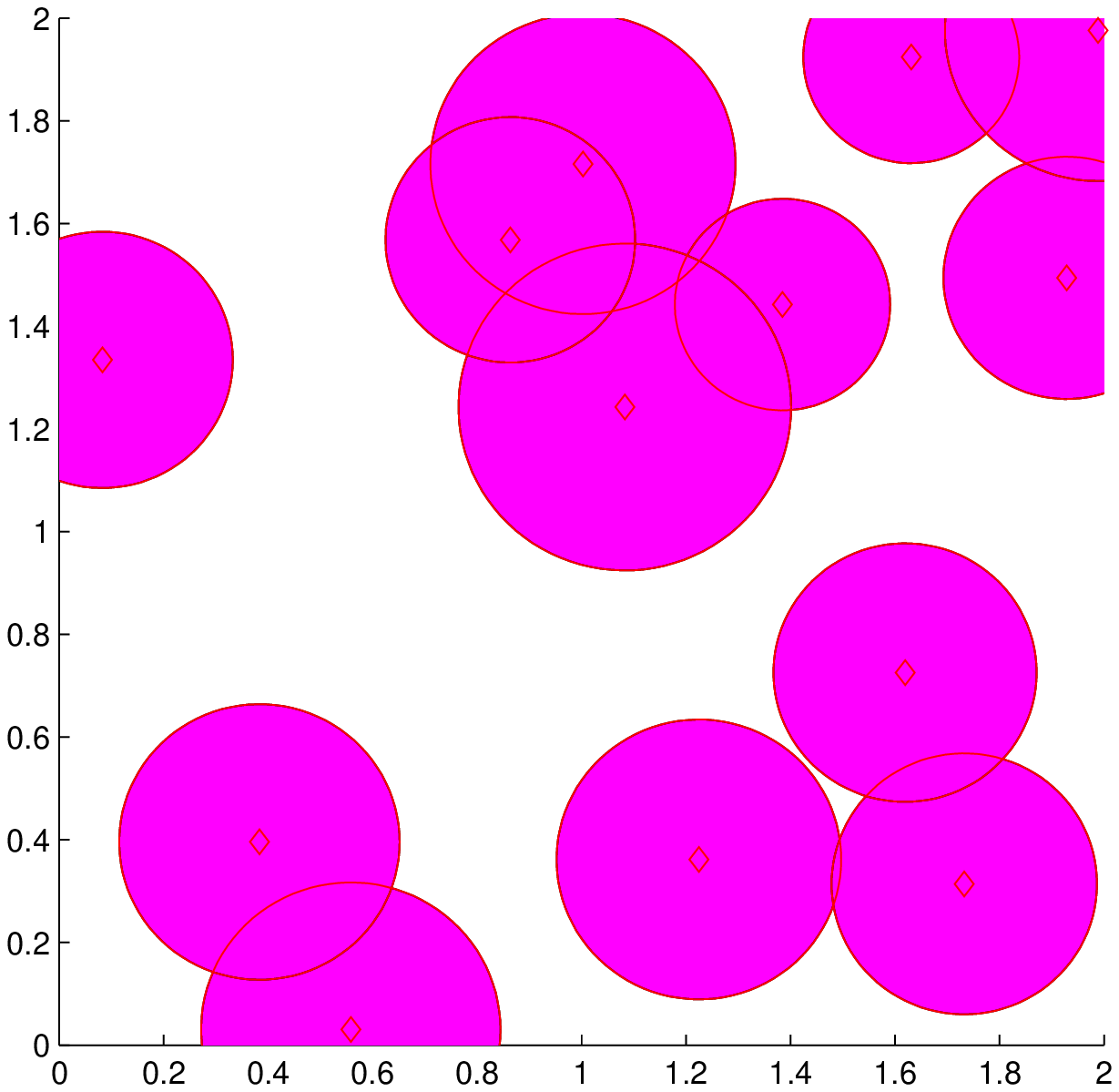}}\\
      \scalebox{0.29}{\includegraphics{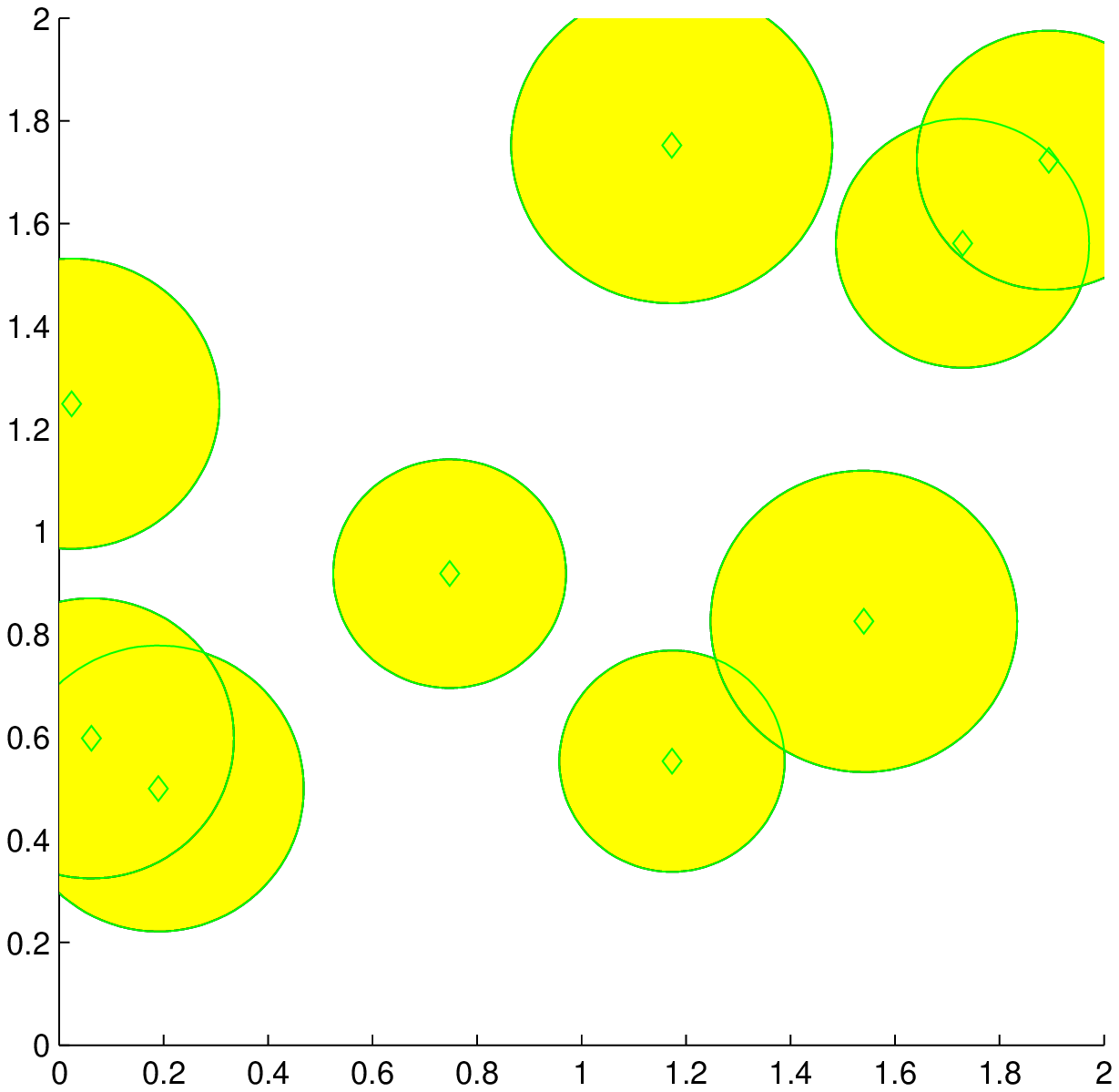}}
      \scalebox{0.29}{\includegraphics{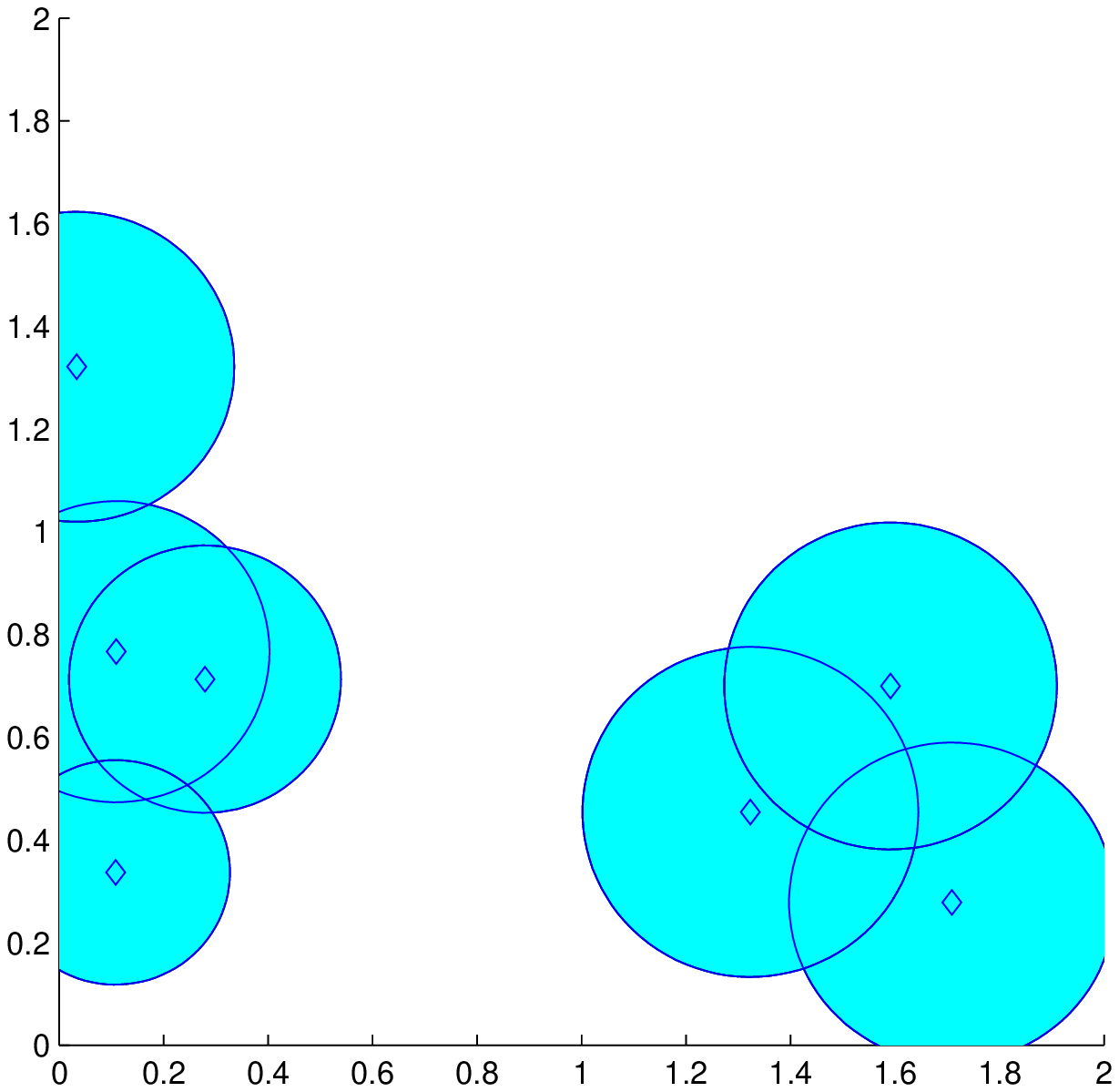}}
\caption{Coverage for each frequency.}
    \label{fig_freq}
  \end{figure}

The frequency planning scheme obtained by our
algorithm for the configuration of Fig. \ref{fig_celco} is
represented in Fig. \ref{fig_planning} on the right. A different color
represent a different frequency. We can see 
that our algorithm has planned four frequencies (black, red, green and
blue) of which we can see the communication area  for each one in
Fig. \ref{fig_freq}.

We give in Algorithm \ref{alg_fapa} the full frequency auto-planning
algorithm. It requires the set of nodes $\omega$, and the neighbor
lists $L_n(v)$ for each node $v$ in $\omega$ to build the abstract
simplicial complex.
If instead of the neighbor lists, one considers the communication
radii $r$, then the abstract simplicial complex is the Vietoris-Rips complex
$\mathcal{R}_{2r}(\omega)$. 
Plus, we need to build the interference graph, so we consider an
interference list $L_i(v)$ for each node $v$ that contains the list of
nodes $v$ has interference with.
Then the algorithm returns the list of assigned
frequencies for every node of $\omega$. 

We introduce three
parameters in the algorithm description that we define here for a
better understanding. First, $N_{\text{assigned}}$ is the number of
nodes to which a frequency is assigned.
Then, $\mathcal{I}$ is a binary
number that is equal to one if there are at least one node with
potentially the same frequency than another node in its interference list
and zero otherwise. Finally, $f(v)$ is the frequency assigned to
the vertex $v \in \omega$. 

For the simplicial complex $X$, we denote $s_k$ the number of its
$k$-simplices, and $v_1,\dots,v_{s_0}$ its $s_0$ vertices.
We do not detail the computations of the degrees $D_1(X),\dots,D_{s_{2}}(X)$
of the $2$-simplices and the indices $I[v_1(X)],\dots,I[v_{s_0}(X)]$
which are explained in \cite{infocom}.
\begin{algorithm}[h]
  \caption{Frequency auto-planning algorithm}
  \label{alg_fapa}
  \begin{algorithmic}[h]
\REQUIRE Set $\omega$ of $N$ vertices, for each vertex $v$ its 
neighbor list $L_n(v)$, and its interference list $L_i(v)$.
\STATE Computation of the abstract simplicial complex $X$ based on
$\omega$ and the $L_n$ lists\;
\STATE Computation of $D_1(X),\dots,D_{s_{2}}(X)$\; 
\STATE Computation of $I[v_1(X)],\dots,I[v_{s_0}(X)]$\; 
\STATE $I_{\max}=\max \{I[v_1(X)],\dots,I[v_{s_0}(X)]\}$\; 
\STATE $N_{\text{assigned}}=0$\;
\STATE $\mathcal{I}=1$\;
\STATE $X'=X$\;
\STATE $i=0$\;
\WHILE{$N_{\text{assigned}} < N$}
\WHILE {$\mathcal{I}==1$} 
\STATE Draw $w$ a vertex of index $I_{\max}$\;
\STATE $X'=X'\backslash \{w\}$\; 
\STATE Re-computation of $D_1(X'),\dots,D_{s'_{2}}(X')$\; 
\STATE Re-computation of $I[v_1(X')],\dots,I[v_{s'_0}(X')]$\; 
 \STATE $I_{\max}=\max \{I[v_1(X')],\dots,I[v_{s'_0}(X')]\}$\;
\STATE $\mathcal{I}=0$\;
\FORALL {$u, v \in X'$}
\IF {$u \in L_i(v)$ \OR $v \in L_i(u)$}
\STATE $\mathcal{I}=1$\;
\ENDIF
\ENDFOR
\ENDWHILE
\FORALL {$v \in X'$}
   \STATE $f(v)=i$\;
\STATE $N_{\text{assigned}}=N_{\text{assigned}}+1$\;
\ENDFOR
\STATE $X'=X \backslash X'$\;
\STATE $i=i+1$\;
\ENDWHILE
\RETURN Assigned frequencies $f(v), \forall v
\in \omega$.
  \end{algorithmic}
\end{algorithm}

\subsection{Simulation and complexity}
First, in simulations, we create the set of nodes with a Poisson
point process of intensity $\lambda=12$ on a square of side $a=2$. 
Then we draw a
communication radius for every node uniformly between $a/10$ and
$2/\sqrt{\pi \lambda}$ and we build the Vietoris-Rips
complex. The complexity of building the complex 
is in $O(N^C)$, where $N$ is the number of nodes and $C$ is the
clique number, i.e. the size of the largest simplex in the complex. If
communication radii are equal to a given $r$, then $C$ is upper-bounded
by $N\left( \frac{r}{a} \right) ^2$ in the general case. This bound
can be improved depending on the percolation regimes, see
\cite{hauteur}. We compute here worst-case complexities in order to
have upper bounds for the complexity of building a simplicial complex,
then for the complexity of our algorithms. In large scale networks,
the simplicial complex representing the network can be build and stored once at the
deployment, then the addition or deletion of some nodes can be done
without re-building the complex entirely, with less complexity.

We want to create distance-based interference
depending on the different communication radii of the nodes, so we
introduce a so-called ``rejection'' radius. Then, each node has a
rejection radius that is a growing function of its communication
radius. In simulations, the rejection radii are equal to half their
corresponding communication radii. If a given
node is inside the disk defined by the rejection radius of another node, then it appears in
its interference list and they are
linked by an edge in the interference graph. The complexity of
building the interference graph is negligible compared to the
complexity of building the abstract simplicial complex.

Then, the complexity of the frequency auto-planning algorithm is
upper-bounded by the complexity of the reduction algorithm that is
called $N_f$ times, if $N_f$ is the number of assigned frequencies.
The complexity of the reduction algorithm is 
$N s_2 \left(N+\sum_{k=3}^{C-1} s_k\right)$ in the general
case according to \cite{infocom}. However it can be simplified when
the set of nodes is drawn uniformly, as in a Poisson point process, on a square, the
communication radii are all equal to a given $r$, and the complex is a
Vietoris-Rips one. In this case, the complexity is in
$O((1+(\frac{r}{a})^2)^{N})$, see \cite{infocom}. 
Therefore the final complexity of our algorithm is in
$O(N_f(1+(\frac{r}{a})^2)^{N})$. 

We can see that the complexity of our algorithm is dependent highly on
the choice of the simplicial complex representation. But this
representation is needed if one wants to compute the topology of a
network, and obtain clustering information without location
information. We do not compare the complexity of our homology-based
algorithm to classic graph-based 
algorithms since the homology parameters we encounter such as the
radius $r$ are not relevant in a graph-based approach.

\subsection{Performance comparison}
In this section, we compare the performance of our frequency
auto-planning algorithm to the greedy coloring algorithm. We choose
this algorithm because our algorithm, as well as the reduction
algorithm, and consequently every algorithm proposed in this article is of greedy
type. Thus we compare two greedy algorithms, ours is simplicial
homology-based, while the coloring algorithm is graph-based.

The frequency planning can be viewed as a graph coloring
problem. If one considers the interference graph, then the optimal
number of frequencies to assign is the chromatic number of the
interference graph. The greedy coloring algorithm provides a coloring
assigning the first new color available for each node. Therefore,
the greedy coloring algorithm is a frequency planning algorithm. And
the number of frequencies planned is at most the maximum node
degree of the interference graph plus one. The greedy coloring gives
especially  good results for sparse graphs as the interference graph is.
So the first parameter that we will use to compare the performance of
both algorithm is the number of planned frequencies.

For each realization of the Poisson process, we
compute the number of frequencies planned by the greedy coloring
algorithm that we denote $N_g$. Then on all realizations with a given
$N_g$, we compute the mean number of frequencies, denoted $N_f$,
planned by our algorithm, so we can see the difference between the two
algorithms. We also indicate which percentage of the
simulations these scenarios are, the occurrence is added for statistical
information. The results are obtained in mean over
$10^4$ configurations. 

\begin{table}[h]
\centering
\begin{tabular}{|cccccc|}
\hline
$N_g$&$2$&$3$&$4$&$5$&$6$\\ \hline
$\E{N_f|N_g}$&$4.00$&$5.04$&$5.83$&$6.43$&$7.13$\\ \hline
Occurrence&$8.8\%$&$55.6\%$&$29.4\%$&$5.3\%$&$0.7\%$\\ \hline
\end{tabular}
\caption{Mean number of planned frequencies.}
\label{table_freq}
 \end{table}

In Table \ref{table_freq}, we can see the mean number of 
frequencies planned by our algorithm given the number of frequencies
planned by the greedy coloring algorithm. We can see that there is a
difference between the two solutions: it is not negligible in the beginning, but it 
decreases with the number of frequencies. Thus, our algorithm
reaches its optimal performance when the number of frequencies grows
for the same mean number of nodes, that is to say when there are
clusters of nodes.

However, even if the greedy coloring algorithm fares well in the
number of planned frequency, it leads to a disparate utilization of
frequencies.  Indeed, if there is only one clique of maximum size, one
frequency will be only used for one node of this clique, and for no
other node in the whole configuration. Therefore, the greedy
coloring algorithm give good results for a homogeneous network, but
not for a cluster network for example. For an optimal utilization of
frequencies, each frequency should cover the whole area, but it is not
always achievable if there are not enough nodes to cover several times the
whole area. Our algorithm aims at a more homogeneous utilization of each ressource.
In order to show that, we compare the percentage of area covered by each frequency
on the total covered area for our frequency auto-planning
algorithm and for the greedy coloring algorithm. We consider an area
to be covered by a frequency if it is inside the communication disk of
a node using this frequency. The percentages are given in mean
over $10^4$ simulations using the same setting as in the first
comparison.

\begin{table}[h]
\hspace{+1cm}
\raggedright
Greedy coloring algorithm\\
\hspace{+1cm}
\begin{tabular}{|ccccc|}
\hline
$N_g$&$3$&$4$&$5$&$6$\\ \hline
$f_1$&$97.3\%$&$97.0\%$&$96.7\%$&$96.3\%$\\ \hline
$f_2$&$47.6\%$&$50.1\%$&$49.9\%$&$51.1\%$\\ \hline
$f_3$&$12.9\%$&$17.7\%$&$18.8\%$&$20.2\%$\\ \hline
$f_4$&&$7.2\%$&$8.8\%$&$8.5\%$\\ \hline
$f_5$&&&$6.2\%$&$6.7\%$\\ \hline
$f_6$&&&&$6.2\%$\\ \hline
Occurrence &$55.6\%$&$29.4\%$&$5.3\%$&$0.7\%$\\ \hline
\end{tabular}
\newline
\newline

\hspace{+1cm} 
Frequency auto-planning algorithm\\
\hspace{+1cm}
\begin{tabular}{|ccccc|}
\hline
$N_f$&$3$&$4$&$5$&$6$\\ \hline
$f_1$&$71.3\%$&$62.1\%$&$54.2\%$&$49.2\%$\\ \hline
$f_2$&$62.3\%$&$56.4\%$&$51.7\%$&$47.4\%$\\ \hline
$f_3$&$35.6\%$&$46.4\%$&$45.7\%$&$42.9\%$\\ \hline
$f_4$&&$24.3\%$&$35.6\%$&$37.8\%$\\ \hline
$f_5$&&&$18.8\%$&$28.1\%$\\ \hline
$f_6$&&&&$15.3\%$\\ \hline
Occurrence &$7.0\%$&$23.0\%$&$29.4\%$&$22.6\%$\\ \hline
\end{tabular}
\caption{Mean percentage of covered area for each frequency}
\label{table_percent}
 \end{table}

We can see in Table \ref{table_percent} the percentage of area covered by each
frequency planned by our algorithm and the greedy coloring
algorithm. The results are presented 
depending on the number of planned frequencies, we also indicate the
number of simulations these results concern for statistical relevance.
For our algorithm, even if the percentage decreases with the order in which the
frequencies are planned, which is logical since the first frequency is
planned in first and so on, we can see that a rather
homogeneous coverage is provided. Doing that, our algorithm maximizes
the usage of each resource. We can see that for the greedy coloring
algorithm, the frequencies are not used equally: the first two
frequencies are always a lot more planned than the other ones, the
latter are thus under-used.

\section{Self-optimization energy conservation algorithm}
\label{sec_opt}
\subsection{Problem formulation}
We are now interested in the self-optimization of a cellular network
previously configured. Indeed, during off-peak hours a cellular
network is under-used, we propose an algorithm that aims at reducing
the energy consumption when user traffic is reduced.
First we want to represent the considered cellular network and its
topology  with an abstract simplicial complex. The network is constituted of
transmitting nodes and their associated coverage disks. In future
cellular networks, these coverage disks can vary in size by modifying
the configuration parameters of the base stations. We choose to
consider here the maximum size of these disks in order to maximize the
coverage of each node so that a maximum number of nodes can be
switched-off. Maximizing the coverage disk of one node induces maximizing the
size of its neighbor list. Then we build the abstract simplicial complex
representing the network and its maximum neighbor lists.
After that, we have to consider boundary nodes
differently from the other ones. Indeed, these nodes allow us to
delimit the area to cover. 

Finally we have to define how we represent user traffic in the
cellular network. In our simulations, we choose to create groups of
network nodes. Then for each group of nodes, the required quality of
service (QoS) is a given number of nodes that are required to stay on
in order to satisfy the traffic of the users inside that group coverage.
Therefore, every group of nodes has a required minimum size.
This QoS metric is quite artificial, but our
algorithm can take into account any QoS as long as it is defined in
terms of required number of ressources from a given pool of ressources
available to a given group of nodes. We choose to implement this simplistic metric
because it is the easiest one to implement. Other metrics can be
considered, the only rule being that one node must pertain to exactly
one group.

\subsection{Algorithm description}
\begin{figure}[h]
      \scalebox{0.29}{\includegraphics{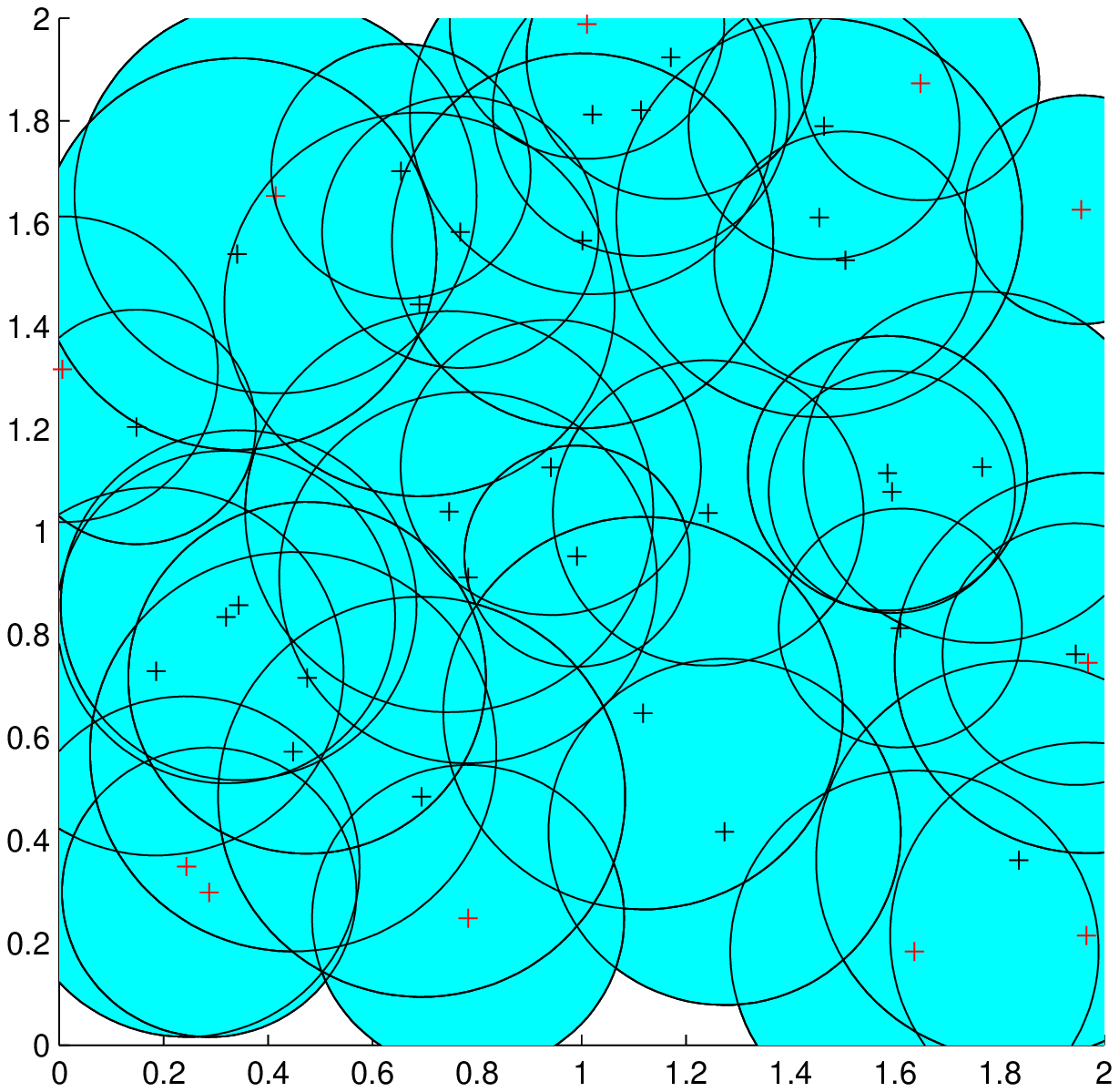}} 
\hfill
      \scalebox{0.29}{\includegraphics{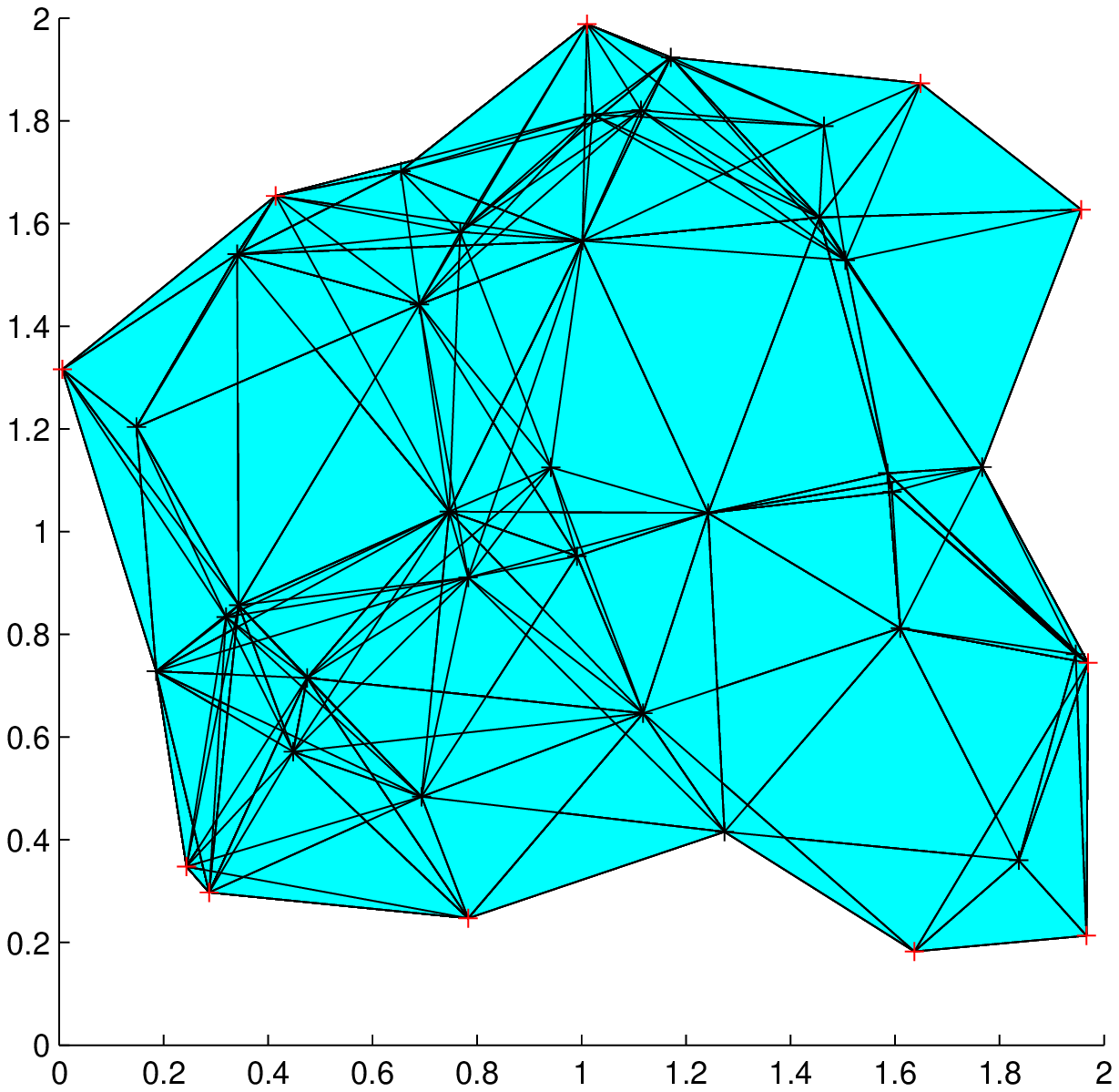}}
\caption{A cellular network and its abstract simplicial complex representation.}
    \label{fig_celcov}
  \end{figure}

We consider a cellular network with transmitting nodes and their
maximal coverage radii that we represent with an abstract simplicial
complex. We can see an example of network and its
complex associated in Fig. \ref{fig_celcov}. The boundary
nodes are computed here via the convex hull and are in red in the figure.

\begin{figure}[h]
\centering
      \scalebox{0.29}{\includegraphics{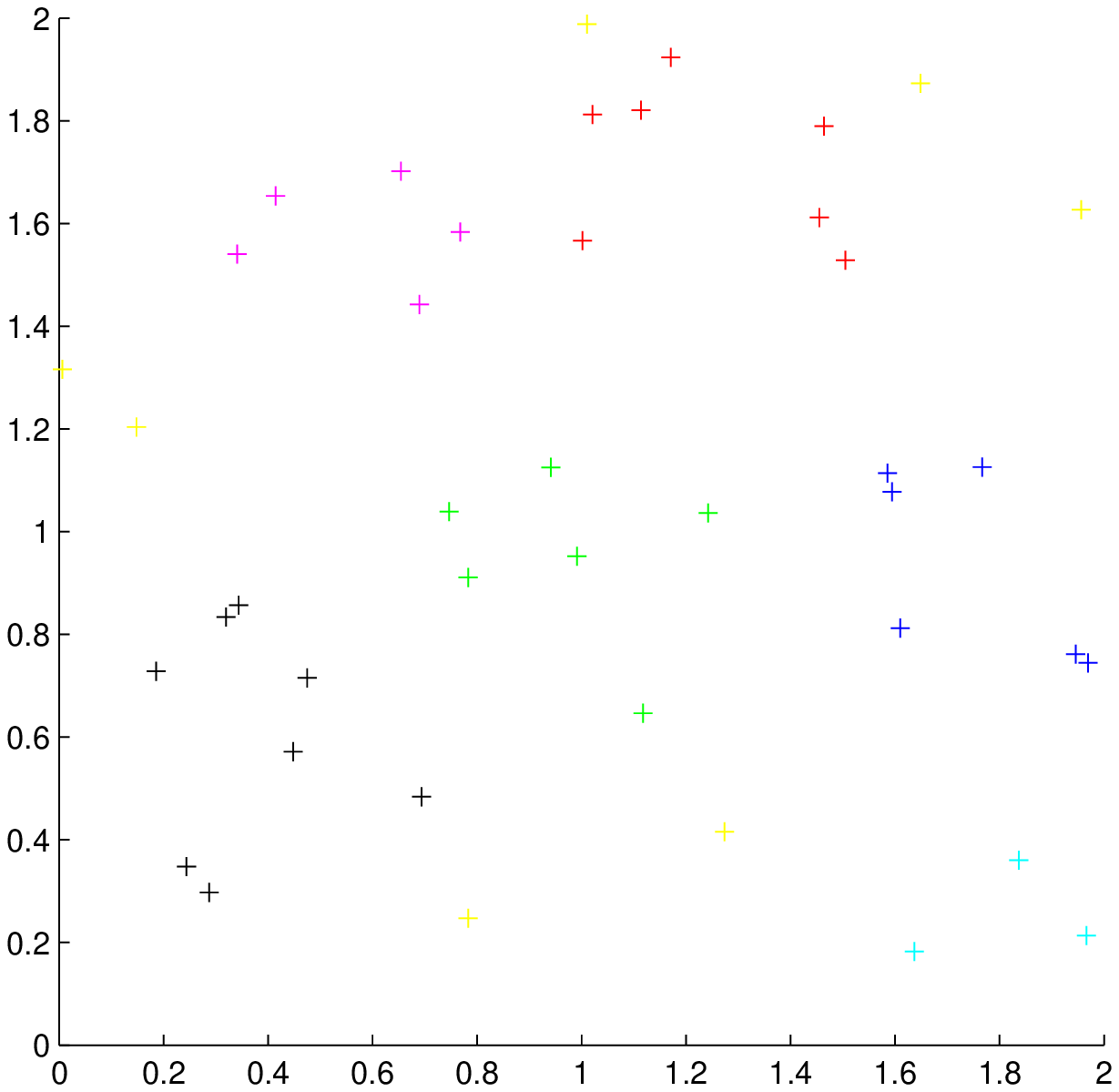}} 
\hfill
\scalebox{0.29}{\includegraphics{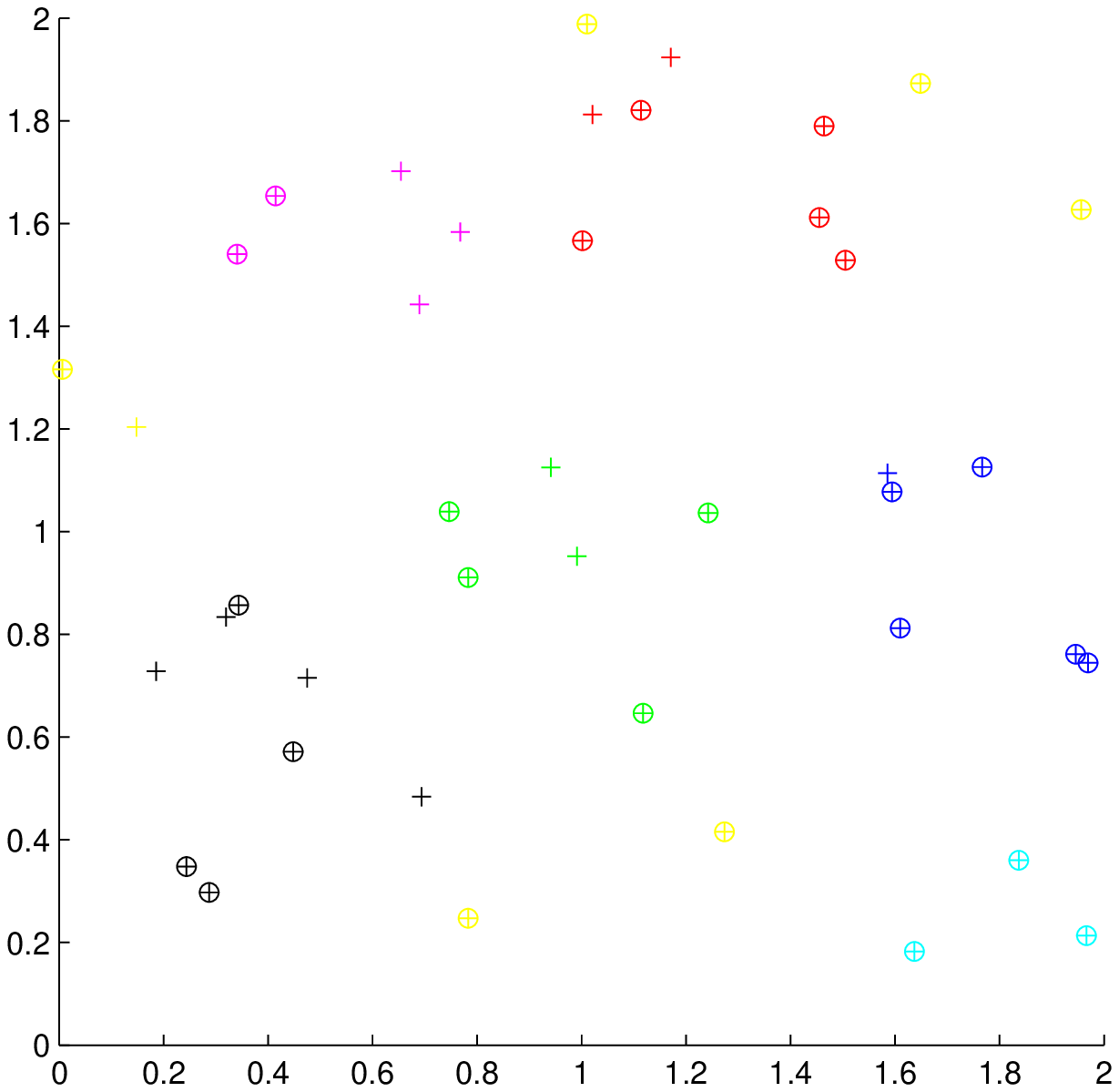}} \\
      \begin{tabular}{|ccccccccccc|}
\hline
Groups&1&\textcolor{red}{2}&\textcolor{blue}{3}&\textcolor{green}{4}
&\textcolor{magenta}{5}&\textcolor{cyan}{6}&\textcolor{yellow}{7}
&\textcolor{yellow}{8}&\textcolor{yellow}{9}&\textcolor{yellow}{10}\\ \hline
Size& $8$ & $7$ & $6$ & $6$ & $5$& $3$&$2$&$2$&$2$&$1$\\ \hline
QoS& $3$ & $5$ & $5$ & $4$ & $2$ &$1$&$1$&$1$&$2$&$1$\\ \hline
\end{tabular}
\caption{QoS groups and required QoS.}
    \label{fig_group}
  \end{figure}

Then for the configuration of Fig. \ref{fig_celcov}, we represent the
QoS groups of nodes in different colors, and give a table with
their corresponding size and required QoS 
in Fig. \ref{fig_group} in the figure on the left. 

The algorithm then begins by the computation of the Betti
numbers. They characterize the topology of the network that we do not
want to modify: the number of connected components and the number of
coverage holes. In the example of Fig. \ref{fig_celcov}, there is one
connected component,$\beta_0=1$, and no coverage hole,
$\beta_1=0$. Then the algorithm aims at reducing the number of nodes
as the reduction algorithm presented in Section \ref{sec_prel}, but we
want to keep enough nodes to satisfy the traffic represented by a
minimum number of nodes to keep on by group. So we apply a modified
reduction algorithm with a different breaking point.
Instead of stopping when the
area is covered by a minimum number of nodes, the algorithm stops
when each group of nodes has been reduced to the size of its
required QoS. We can see the obtained result for the configuration of
Fig. \ref{fig_celcov} in Figure
\ref{fig_group} on the right. The kept nodes are circled.

\begin{figure}[h]
      \scalebox{0.29}{\includegraphics{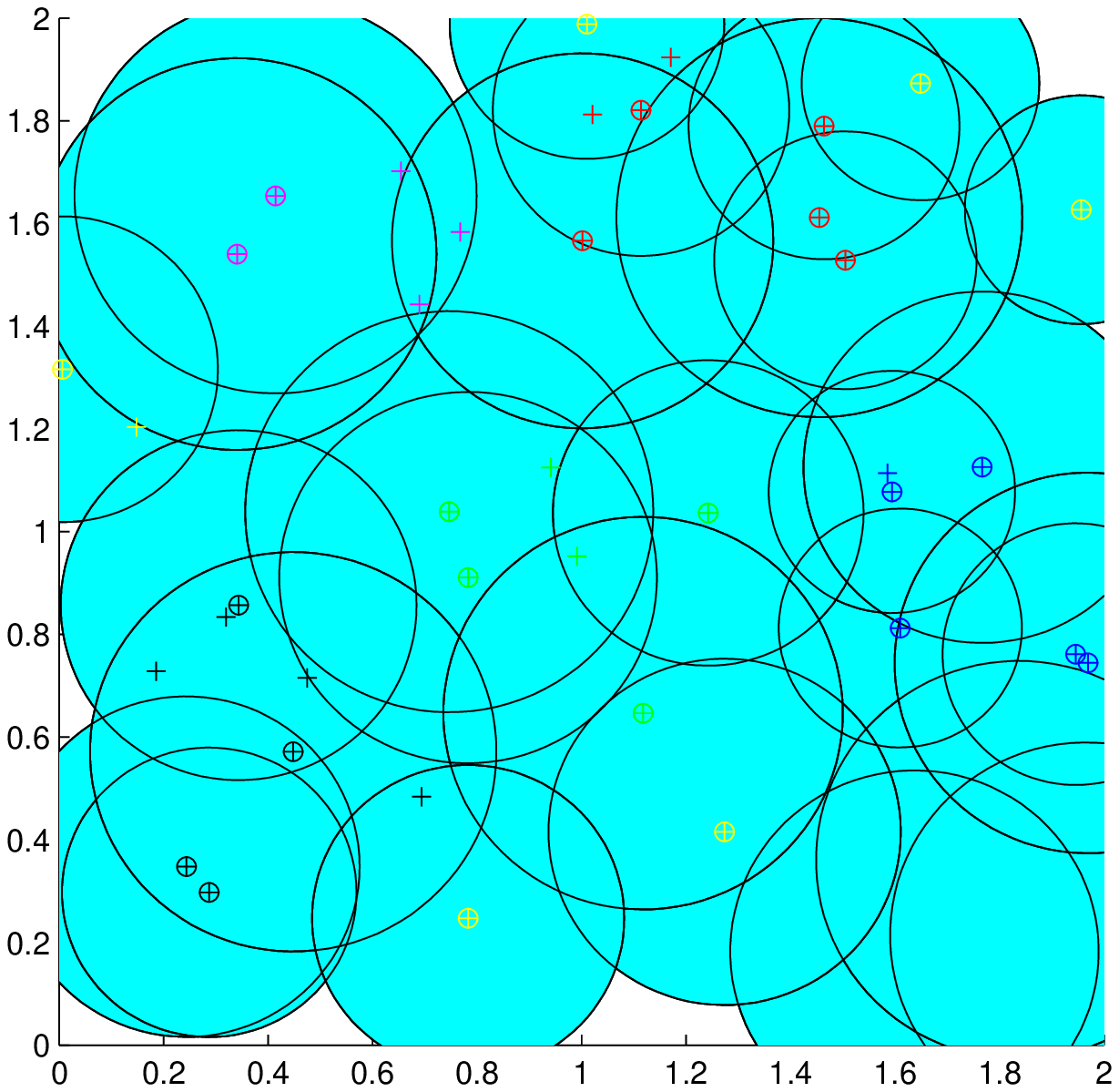}} 
\hfill
      \scalebox{0.29}{\includegraphics{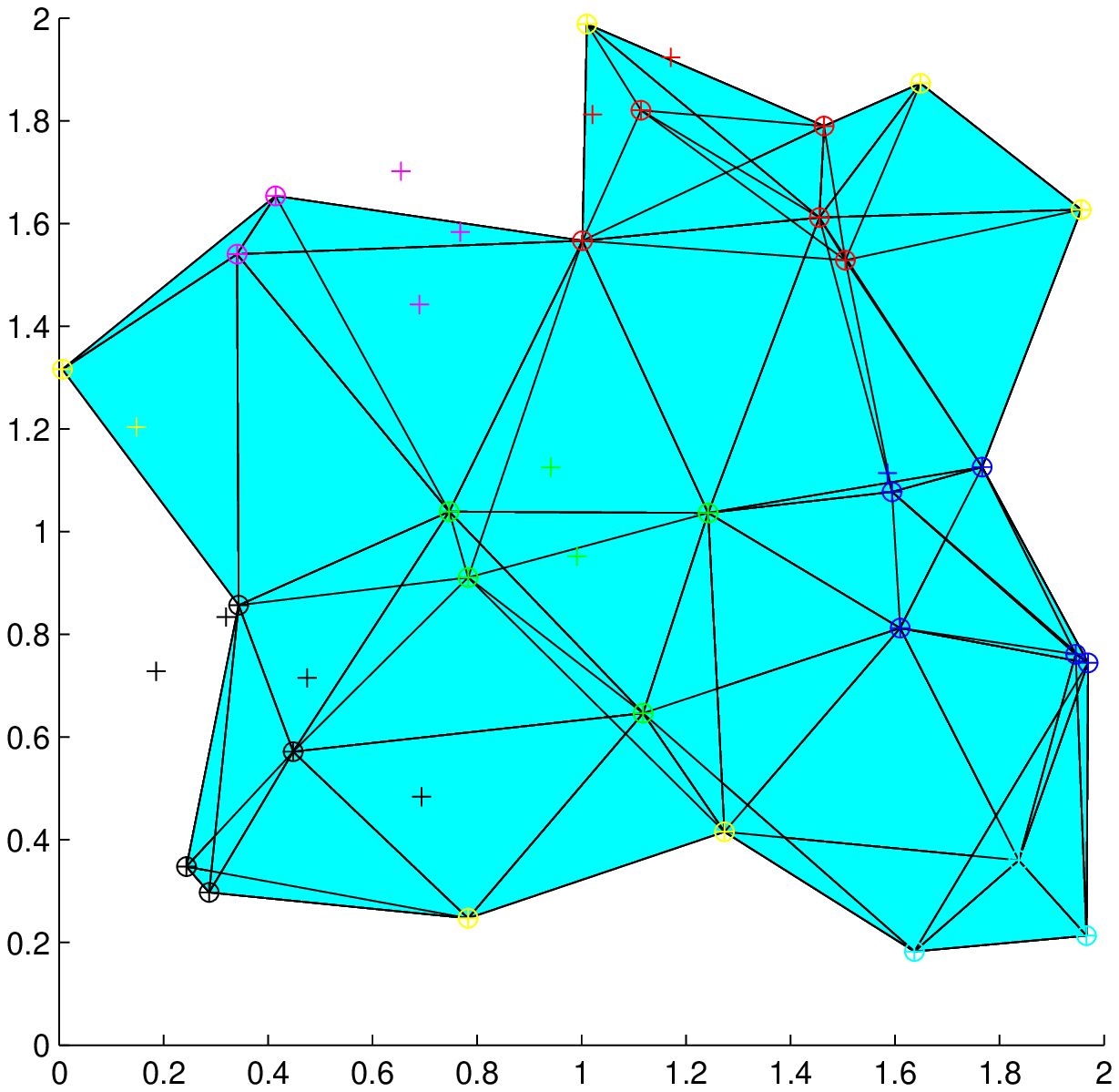}}
\caption{The network after the first step.}
    \label{fig_step1}
  \end{figure}

The cellular network and its abstract simplicial complex representation is
represented in Fig. \ref{fig_step1}.

\begin{figure}[h]
      \scalebox{0.29}{\includegraphics{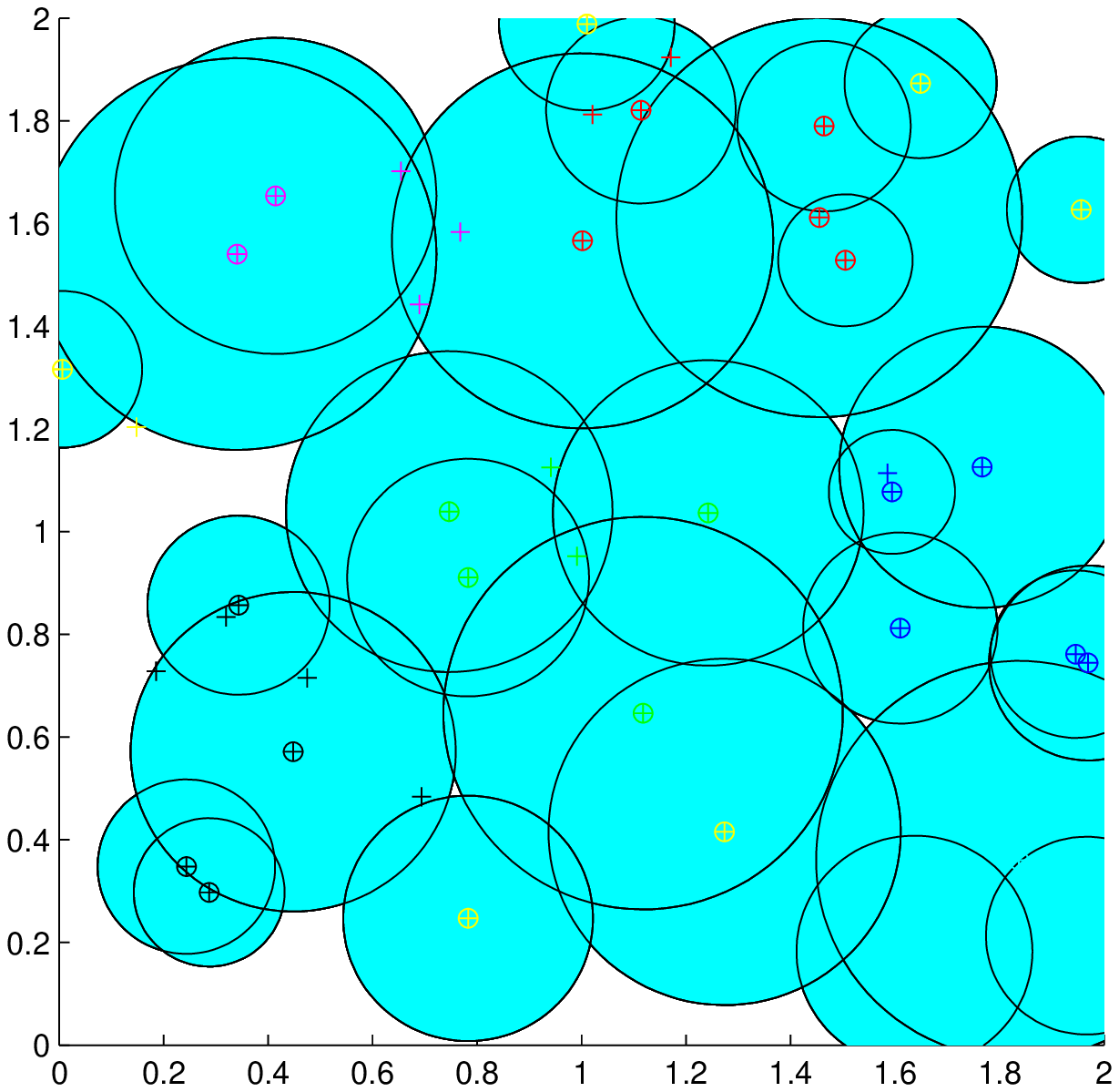}} 
\hfill
      \scalebox{0.29}{\includegraphics{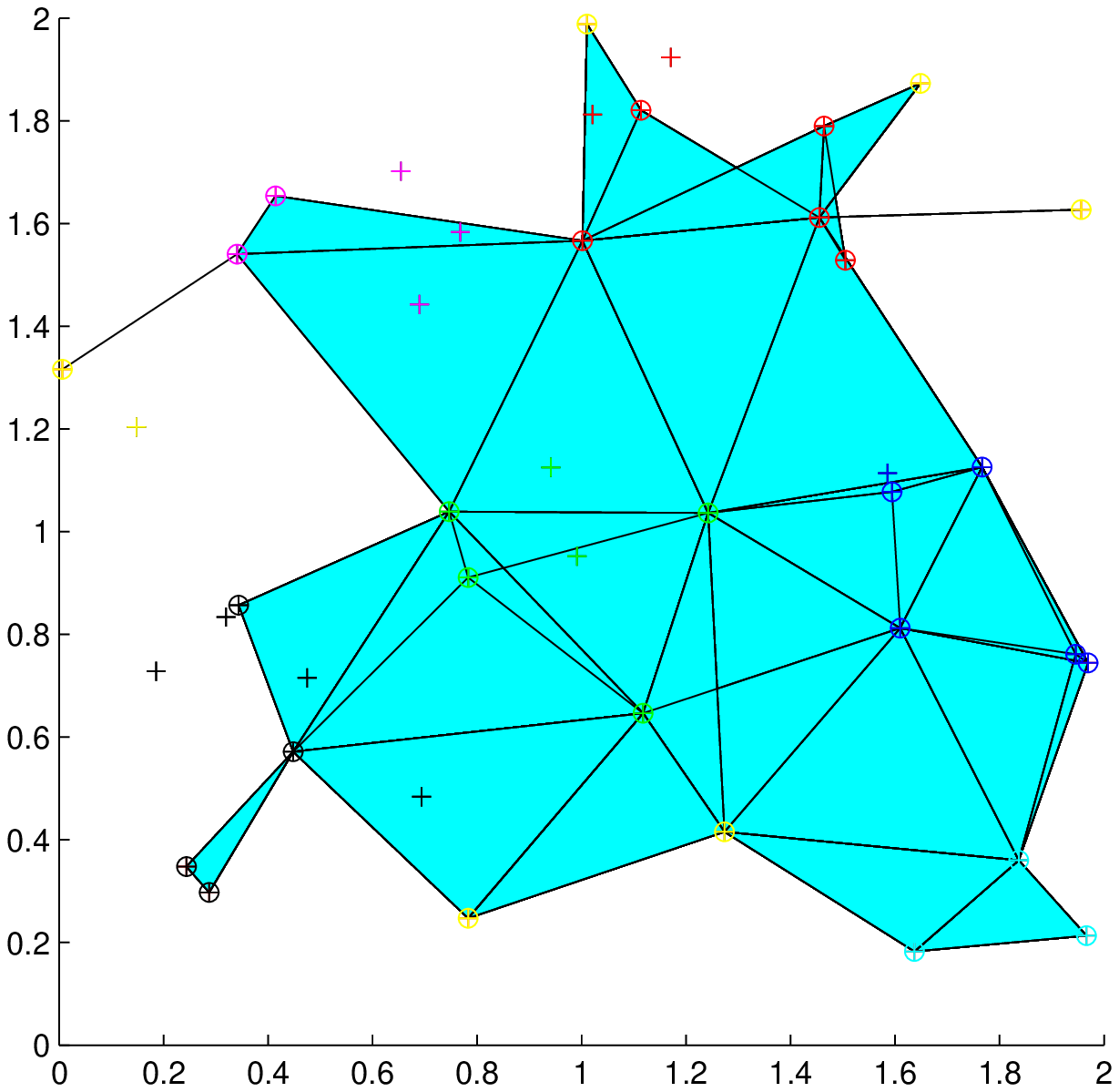}}
\caption{Final configuration.}
    \label{fig_res}
  \end{figure}

Finally, given this configuration of switched-on nodes, the algorithm
tries to reduce as much as possible the 
coverage radii of each node without creating a coverage hole to
minimize the energy consumption. 
In Fig. \ref{fig_res}, we can see the final configuration of
the cellular network with the optimized coverage radii and its
abstract simplicial complex representation for the 
configuration of Fig. \ref{fig_celcov}.

In the end, we have a configuration of
nodes to keep switched-on that is optimal. From the first part of
the algorithm, we ensure that enough nodes are kept on to satisfy
the QoS representing user traffic. Then, in the second part of the
algorithm, we ensure that no energy 
is spend uselessly by optimizing the size of the serving cells.

We give in Algorithm \ref{alg_opt} the full energy conservation
algorithm. It requires the set of nodes $\omega$ and their
neighbor lists in order to build the abstract simplicial complex.
If instead of the neighbor lists, one considers the communication
radii $r$, then the abstract simplicial complex is the Vietoris-Rips complex
$\mathcal{R}_{2r}(\omega)$. Or if one wants to represent the exact
topology and considers the coverage
radii $r$, then it is the \u{C}ech complex $\mathcal{C}_{r}(\omega)$.

The groups are represented by the variable $G(v)$ that for each
node $v$ gives its group number. Then for a given group $g$, $Q(g)$ is the minimum
number of nodes required by the QoS, while $S(g)$ is
its number of nodes. So we always have $Q(g)\leq S(g)$. Either these
parameters have to be given as 
input, or the way to compute them is to be integrated into the
algorithm. 
We can see that the breaking point of the ``while'' loop of the
algorithm takes the QoS parameter $Q$ into account. Then a node can be removed
if and only if it does not modify the Betti numbers of the abstract
simplicial complex, and it is not needed for the QoS
requirements. 

Optional coverage radii reduction for each node is done in the last
loop. To implement that last step, maximum coverage radii are needed
in input. The order in which the coverage radii are examined is
random. The reduction can be made by steps for example.
At the end, the algorithm returns the list of kept
nodes with their new coverage radii.
\begin{algorithm}[h]
  \caption{Energy conservation algorithm}
  \label{alg_opt}
  \begin{algorithmic}[h]
    \REQUIRE Set $\omega$ of $N$ vertices, for each vertex $v$ its 
neighbor list $L_n(v)$, its group $G(v)$, and its maximum
coverage radius $r(v)$. For each group $g$, its size $S(g)$
and its QoS $Q(g)$.
\STATE Computation of the abstract simplicial complex $X$ based on
$\omega$ and the $L_n$ lists.\;
\STATE Creation of the list of boundary vertices $L_B$\;
\STATE Computation of $\beta_0(X)$ and $\beta_{1}(X)$\; 
\STATE Computation of $D_1(X),\dots,D_{s_2}(X)$\; 
\STATE Computation of $I[v_1(X)],\dots,I[v_{s_0}(X)]$\; 
\FORALL{$ v \in L_B$} \STATE
    $I[v]=-1$\;
\ENDFOR
 \STATE $I_{\max}=\max \{I[v_1(X)],\dots,I[v_{s_0}(X)]\}$\; 
\WHILE {$I_{\max}> 2 $ \AND $S(g) \geq Q(g)$ $\forall g$} 
\STATE Draw $w$ a vertex of index $I_{\max}$\;
\STATE $X'=X\backslash \{w\}$\; 
\STATE Computation of $\beta_0(X'), \beta_{1}(X')$\; 
\IF {$\beta_0(X')\neq\beta_0(X)$ \OR $\beta_1(X')\neq\beta_1(X) $
  \OR $S(G((w))< Q(G((w))$}
    \STATE $I[w]=-1$\; 
\ELSE{} 
\STATE $S(G(w))=S(G(w))-1$\;
\STATE Re-computation of $D_1(X'),\dots,D_{s'_{2}}(X')$\; 
\STATE Re-computation of $I[v_1(X')],\dots,I[v_{s'_0}(X')]$\; 
    \STATE $I_{\max}=\max \{I[v_1(X')],\dots,I[v_{s'_0}(X')]\}$\;
    \STATE $X=X'$\;
    \ENDIF
    \ENDWHILE
\FORALL {$v \in X$ taken in random order}
\STATE $X'=X$\;
\WHILE{$\beta_0(X')=\beta_0(X)$ \AND$\beta_1(X')=\beta_1(X)$}
\STATE Reduce $r(v)$\;
\ENDWHILE
\STATE $X=X'$\;
\ENDFOR
    \RETURN List of kept vertices and their new coverage radii $r(v)$.
  \end{algorithmic}
\end{algorithm}

\subsection{Simulation and complexity}
As in the previous section, we simulate the set of nodes with a
Poisson point process on a square of side $a=2$. However, since we
need boundary nodes and do not need a great number to see the
efficiency of the algorithm, the intensity of the Poisson point process is
lowered to $\lambda=6$. Then the coverage radii of the vertices from
the Poisson point process are sampled uniformly between $a/10$ and
$2/\sqrt{\pi \lambda}$. And the coverage radii of the boundary nodes
is set to $a/3$. The complexity of building the Vietoris-Rips complex 
is still in $O(N^{N\left( \frac{r}{a} \right) ^2})$, where $N$ is the number of vertices and $r$ is the
common coverage radius. 

To create the groups of nodes in simulations, we have to take some
care. Indeed, these groups have to make sense geographically, but we
do not have location information. So in order to consider clusters of
nodes, every group is defined by a maximum simplex, a simplex which is
not the face of any other simplex. The groups are created so that
every node pertains to exactly one group. Then, for every group $g$ of
size $S(g)=k$, an integer is uniformly drawn between $1$ and $k$, this integer
is then the required number of nodes to keep on, i.e. $Q(g)$.

For our simulations, we choose to give advantage to the
larger simplices for the constitution of the groups. Thus, 
the first group will consist of the largest simplex, or one randomly
chosen among the largest ones, then simplices of smaller size will
become groups until every node is part of a group. It is possible to
consider other rules for the constitution of the groups, but it has to
follow one condition: every node must pertain to exactly one
group. The constitution of $N_G$ groups is given in Algorithm
\ref{alg_qos}. 
\begin{algorithm}[h]
  \caption{Computation of QoS groups}
  \label{alg_qos}
  \begin{algorithmic}[h]
\REQUIRE Simplicial complex $X$
\STATE $N_G=0$\;
\FORALL {Simplex $S_k \in X$ from largest to smallest}
\IF {$\forall v \in S_k, G(v)==0$}
\STATE $N_G=N_G+1$
\STATE $\forall v \in S_k G(v)=N_G$\;
\STATE $S(N_G)=k+1$\;
\STATE Draw $Q(N_G)$ among $\{1,\dots,k+1\}$\;
\ENDIF
\ENDFOR
  \end{algorithmic}
\end{algorithm}

The complexity of the energy conservation algorithm is the same as the
reduction algorithm since it is an improved reduction algorithm, that
is $O((1+(\frac{r}{a})^2)^{N})$, see \cite{infocom}. 

\subsection{Performance comparison}
In this section,
we compare the performance of our algorithm to an optimal, not always
achievable greedy solution. As in the previous section we choose to
compare our greedy simplicial homology-based energy conservation
algorithm to a greedy graph-based algorithm.
We do not know of a energy conservation
algorithm that can switch-off nodes during off-peak hours while
maintaining coverage. Thus, we compare the number of switched-on nodes
after the execution of our energy conservation algorithm, to the number
of nodes needed for the QoS, given by the sum of the $Q(g)$ for all $g$. It is
important to note that this optimal solution is not always achievable
because it does not take into account that the area is to stay
covered. Some nodes have to be kept for traffic reasons, while
other are kept to maintain connectivity and/or coverage. This number
of nodes can not be obtained for a random configuration of nodes
whose positions do not follow a given pattern.

Our simulation results are computed on $10^4$ configurations of 
We denote by $N_o=\sum_g Q(g)$ the optimal
number of kept nodes for traffic reasons, and by $N_k$ the number of kept nodes with our
energy conservation algorithm for both traffic and coverage reasons. 
First we compute the percentage of simulations for which we
have a given difference between 
the obtained number $N_k$ and the minimum number $N_o$ of kept nodes.

\begin{table}[h]
\centering
\begin{tabular}{|cccccc|}
\hline
$N_k-N_o$& $0$ & $1$ & $2$ & $3$&$4$\\ \hline
Occurrence&$1.3\%$&$4.2\%$&$8.5\%$&$12.5\%$&$15.4\%$\\ \hline
\hline
$5$&$6$&$7$&$8$&$9$&$\geq 10$\\ \hline
$15.4\%$&$14.3\%$&$11.0\%$&$7.5\%$&$4.9\%$&$4.7\%$\\ \hline
\end{tabular}
\caption{Occurrences of given differences between $N_k$ and $N_o$.}
\label{table_qos}
 \end{table}

We can see in Table \ref{table_qos} the percentage of simulations in
which the number of kept nodes is different from the optimal
number of nodes. For $1.3\%$ of the simulations the optimal number
is reached. In $82.8\%$ of our simulations the difference between
the optimal and the effective number of kept nodes is smaller than
$7$, and it never exceeds $18$. 
From our simulations, an average of $N_k-N_o=5.16$ nodes
are kept by our algorithm only for coverage reasons, and not for
traffic reasons. To explain this number, we can note that 
the $16$ boundary nodes are always kept for coverage
reasons, even if they can also be used for traffic reasons depending on
the configurations. 

To have more advanced comparison, 
for the $10^4$ configurations, we compute the optimal number of
nodes. Then for each optimal number of nodes $N_o$ that occurred
the most, we compute the mean number of kept nodes $\E{N_k|N_o}$
over the simulations which have $N_o$ for optimal number. The results
are given in Table \ref{table_qos2}. For comparison, we also compute the difference between $N_o$ and $N_k$ in
percent. We finally indicate which percent of our $10^4$ simulations
these cases occur to show the relevance of these statistical results. 

\begin{table}[h]
\centering
\begin{tabular}{|cccccc|}
\hline
$N_o$&$22$&$23$&$24$&$25$&$26$\\ \hline
$\E{N_k|N_o}$&$28.95$&$29.52$&$30.04$&$30.69$&$31.30$\\ \hline
Difference&$31.6\%$&$28.4\%$&$25.1\%$&$22.8\%$&$20.4\%$\\ \hline
Occurrence&$5.0\%$&$6.2\%$&$7.0\%$&$7.4\%$&$7.9\%$\\ \hline
\hline
$27$&$28$&$29$&$30$&$31$&$32$\\ \hline
$31.85$&$32.68$&$33.25$&$33.95$&$34.64$&$35.38$\\ \hline
$18.0\%$&$16.7\%$&$14.7\%$&$13.1\%$&$11.7\%$&$10.6\%$\\ \hline
$7.4\%$&$7.5\%$&$7.2\%$&$6.3\%$&$5.9\%$&$4.8\%$\\ \hline
\end{tabular}
\caption{Mean number of kept nodes $\E{N_k|N_o}$ given $N_o$.}
\label{table_qos2}
 \end{table}
 
We can see in Table \ref{table_qos2}
that the more nodes are needed, the less difference there is between
the number of kept nodes for both coverage and traffic reasons $N_k$
and the optimal minimum number of nodes kept for only traffic reasons
$N_o$. Indeed, if more nodes are needed for traffic, 
there is a great chance that these nodes can cover the whole area, and
fewer nodes are needed only for coverage reasons.

\section{Disaster recovery algorithm}
\label{sec_pmr}
In this section we present a disaster recovery algorithm introduced in
\cite{pmr} of which we remind the main idea and investigate more
thoroughly the performance in the next section. But first, we need to
introduce some stochastic geometry concepts.

\subsection{Repulsive point processes}
In this section, we aim at repairing a damaged
cellular network by adding new nodes randomly in the damaged area.
The most common point process in
cellular network representation is the Poisson point process. However
in this process, conditionally to the number of points,
their positions are independent from each other. This
independence creates some aggregations of points, as well as voids
(i.e. coverage holes), both of which are not convenient for the
recovery of a network. That is why we propose the use of
determinantal point processes, in which the points positions are not
independent anymore. 

General point processes can be characterized by
their so-called Papangelou intensity. Informally speaking, for $x$ a
location, and $\omega$ a realization of a given point process, that is
a set of points, $c(x,\omega)$ is the probability to have a point
in an infinitesimal region around $x$ knowing the set of points
$\omega$. For Poisson process, $c(x, \omega)=1$ for any $x$ and any
$\omega$. A point process is said to be repulsive (resp. attractive)
whenever  $c(x,\omega) \ge c(x, \zeta)$ (resp. $c(x, \omega) \le c(x,
\zeta)$) as soon as $\omega \subset \zeta$. For repulsive point
process, that means that the greater the set of points, the smaller the probability
to have an other point.

Among repulsive point processes, we are in particular interested in
determinantal processes:
\begin{definition}[Determinantal point process]
Given $X$ a Polish space equipped with the Radon measure $\mu$, and $K$
a measurable complex function on $X^2$, we say that $N$ is a
determinantal point process on $X$ with kernel $K$ if it is a point
process on $X$ with correlation functions $\rho_n(x_1,\dots,x_n)=\det
(K(x_i,x_j)_{1\leq i,j \leq n})$ for every $n\geq 1$ and $x_1,\dots
,x_n \in X$.
\end{definition}
We can see that when two points $x_i$ and $x_j$ tend to be close to
each other for $i\neq j$, the
determinant tends to zero, and so does the correlation function. That
means that the points of $N$ repel each other.
There exist as many determinantal point processes as
functions $K$. We are interested in the following:
\begin{definition}[Ginibre point process]
The Ginibre point process is the determinantal point process with kernel
$K(x,y)=\sum_{k=1}^{\infty}B_k\phi_k(x)\overline {\phi_k(y)}$, where
$B_k, k=1,2,\dots$, are $k$ independent Bernoulli variables and
$\phi_k(x)=\frac{1}{\sqrt{\pi k!}}e^{\frac{-|x|^2}{2}}x^k$ for $x \in
\mathbb{C}$ and $k \in \mathbb{N}$. 
\end{definition}

The Ginibre point process is invariant with respect to
translations and rotations, making it relatively easy to simulate on a
compact set. Moreover, the repulsion induced by a Ginibre point
process is of electrostatic type. The principle behind the repulsion
lies in the probability density used to draw points positions. The
probability to draw a point at the exact same position of an already
drawn point is zero. Then, the probability increases with increasing
distance from every existing points. Therefore the probability to
draw a point is greater in areas the furthest away from every
existing points, that is to say in coverage holes. 
The simulation of Ginibre determinantal point processes is detailed in \cite{ian}.

\subsection{Main idea}
The disaster recovery algorithm aims at restoring a damaged cellular
network. Thus, we consider a cellular network presenting coverage
holes and possibly many disconnected components. The \u{C}ech complex
or its approximation the Vietoris-Rips complex is build based on the
set of nodes and either their coverage or communication radii, or
their neighbor lists. We also need a list of boundary
nodes, which can be fictional, in order to know the whole area to be
covered once the network is repaired. 

\begin{figure}[h]
\centering
      \scalebox{0.29}{\includegraphics{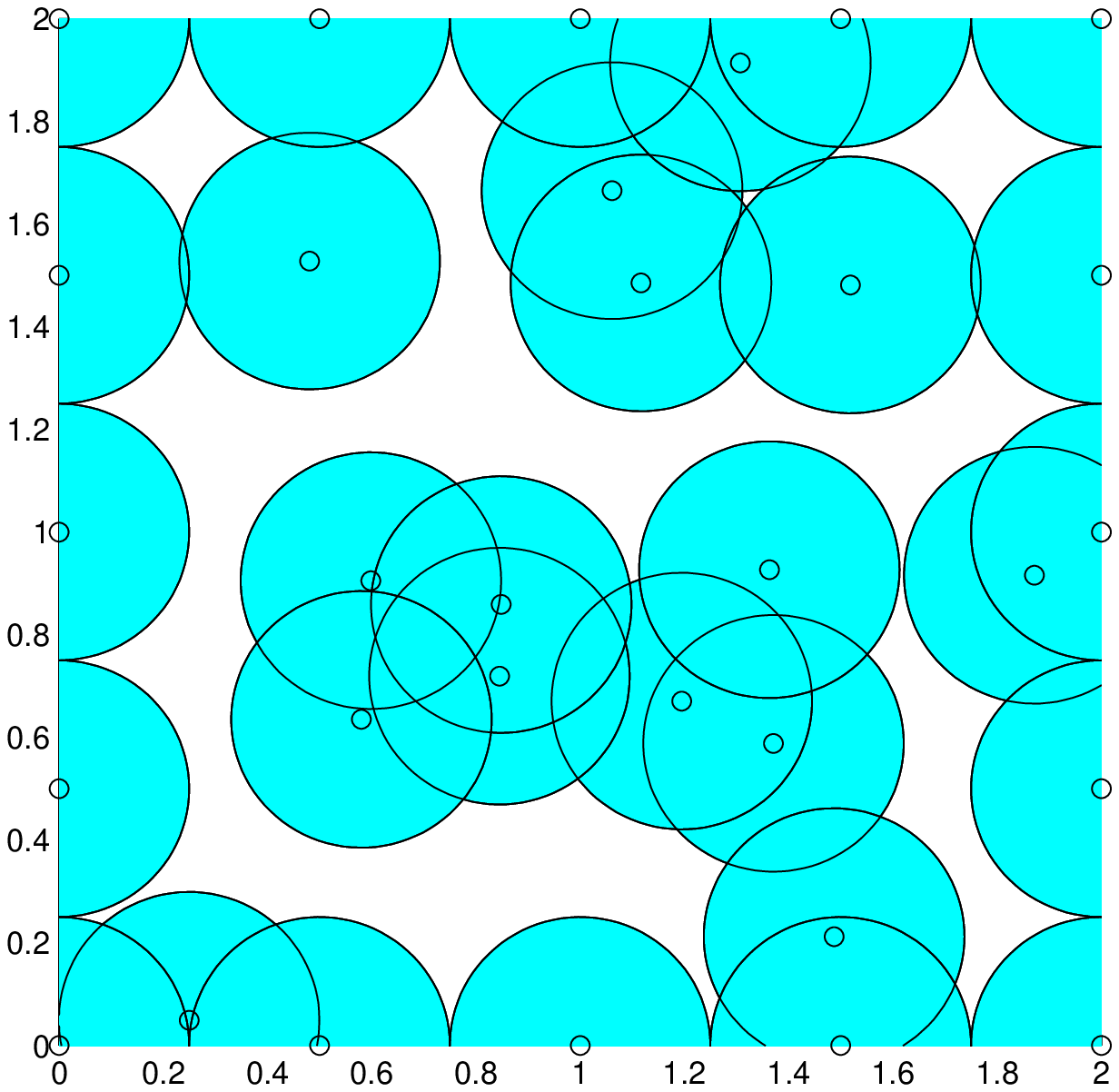}} 
\hfill
\scalebox{0.29}{\includegraphics{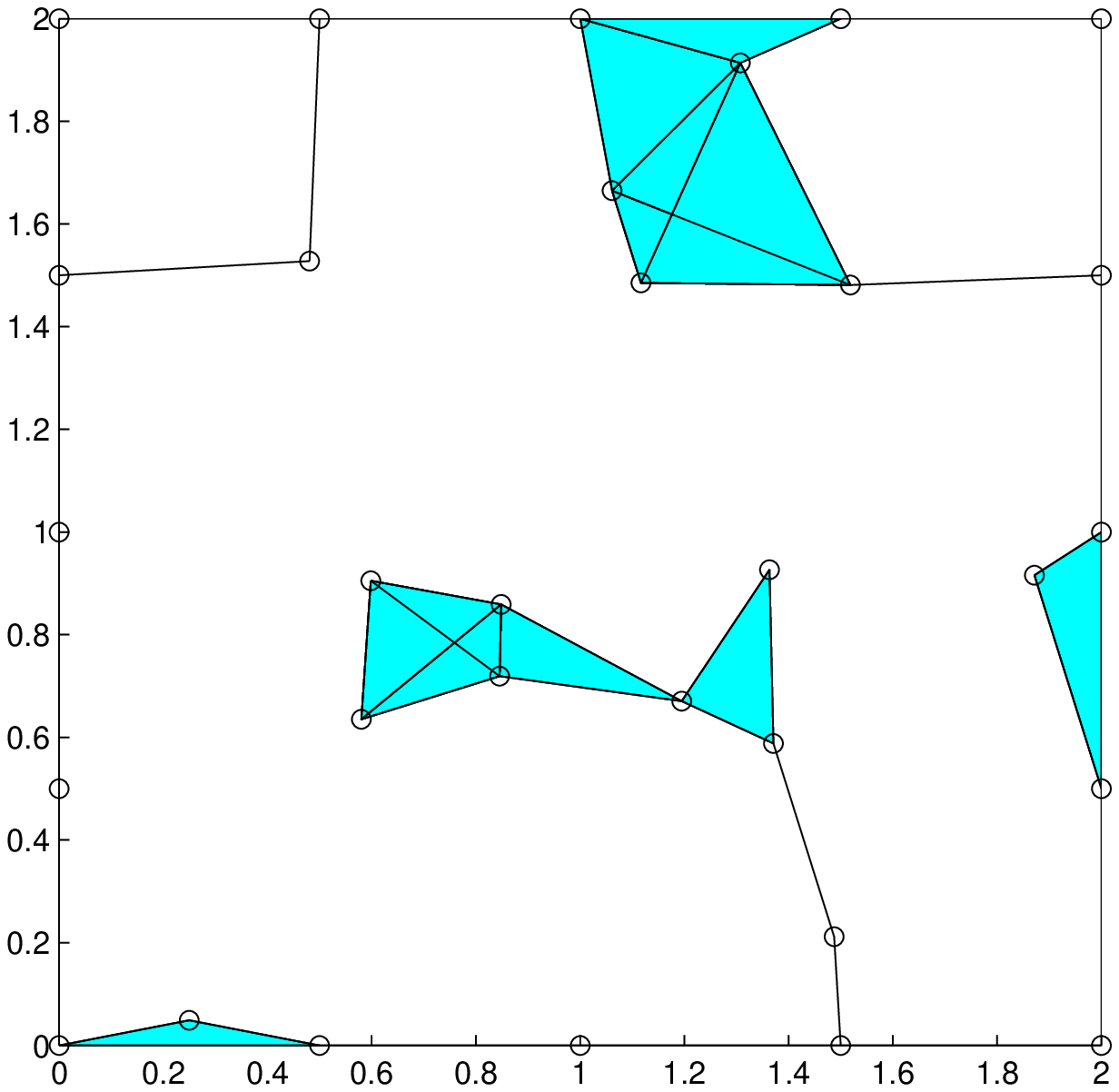}} 
    \caption{A damaged cellular network with a fixed boundary.}
    \label{fig_damaged}
  \end{figure}

In this section we suppose that the node locations are known. Indeed
the disaster recovery algorithm provides the locations where to put
new nodes in order to patch the network. We
also restrict ourselves to a fixed common coverage radius for every
nodes, even if the main idea can be extended. We can
see an example of damaged cellular network with a square boundary of
fictional nodes and its representation by the approximated coverage
complex: the Vietoris-Rips complex in Fig. \ref{fig_damaged}.

The algorithm then adds new nodes in addition to the set of existing
nodes. For the addition of nodes, several methods, deterministic or
random, are possible. But we choose here to focus here on the most
fitted random method: determinantal point processes. However, the
number of new added nodes is not computed as in determinantal point
processes, but via an incrementation. It is first set to the minimum
number of nodes to cover the whole area minus the number of existing
nodes. When the nodes are added, the two first Betti numbers are
computed. Then if there are more than one connected component, or any
coverage hole, other new nodes are added. Their number is incremented
with a random variable following an exponential growth: first set to
$1$, it is doubled every time new nodes are added and the network is
still not patched. 

Using determinantal point processes for the addition of nodes allows
us to not only take into account the number of existing nodes via
the computation of the number of added nodes, but also their
locations. Indeed, the existing nodes are considered part of the point
process, then new nodes positions are drawn following a determinantal
point process. 
The addition of new nodes stops as soon as the network is repaired:
one connected component and no coverage hole.

\begin{figure}[h]
      \scalebox{0.29}{\includegraphics{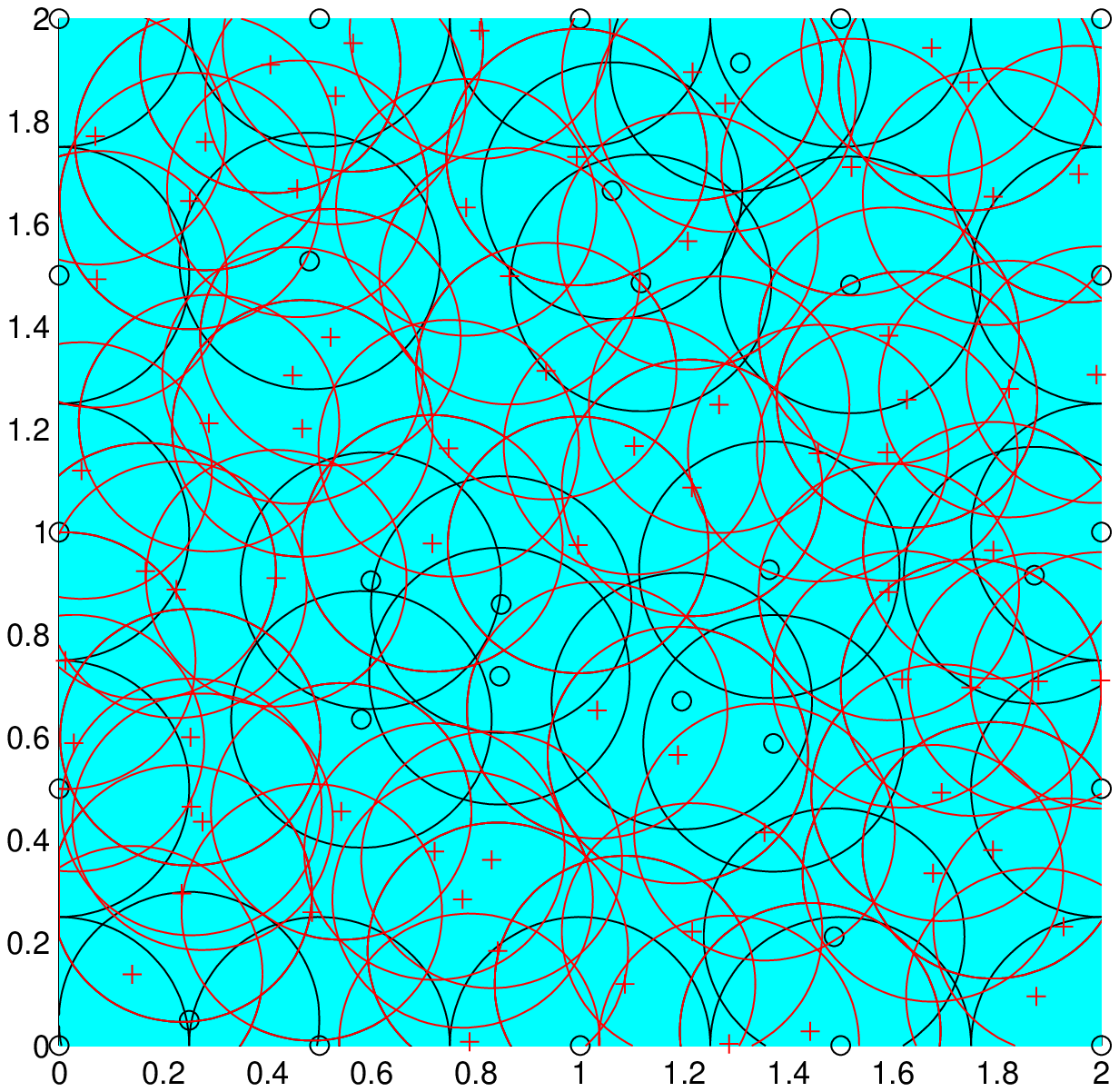}} 
\hfill
      \scalebox{0.29}{\includegraphics{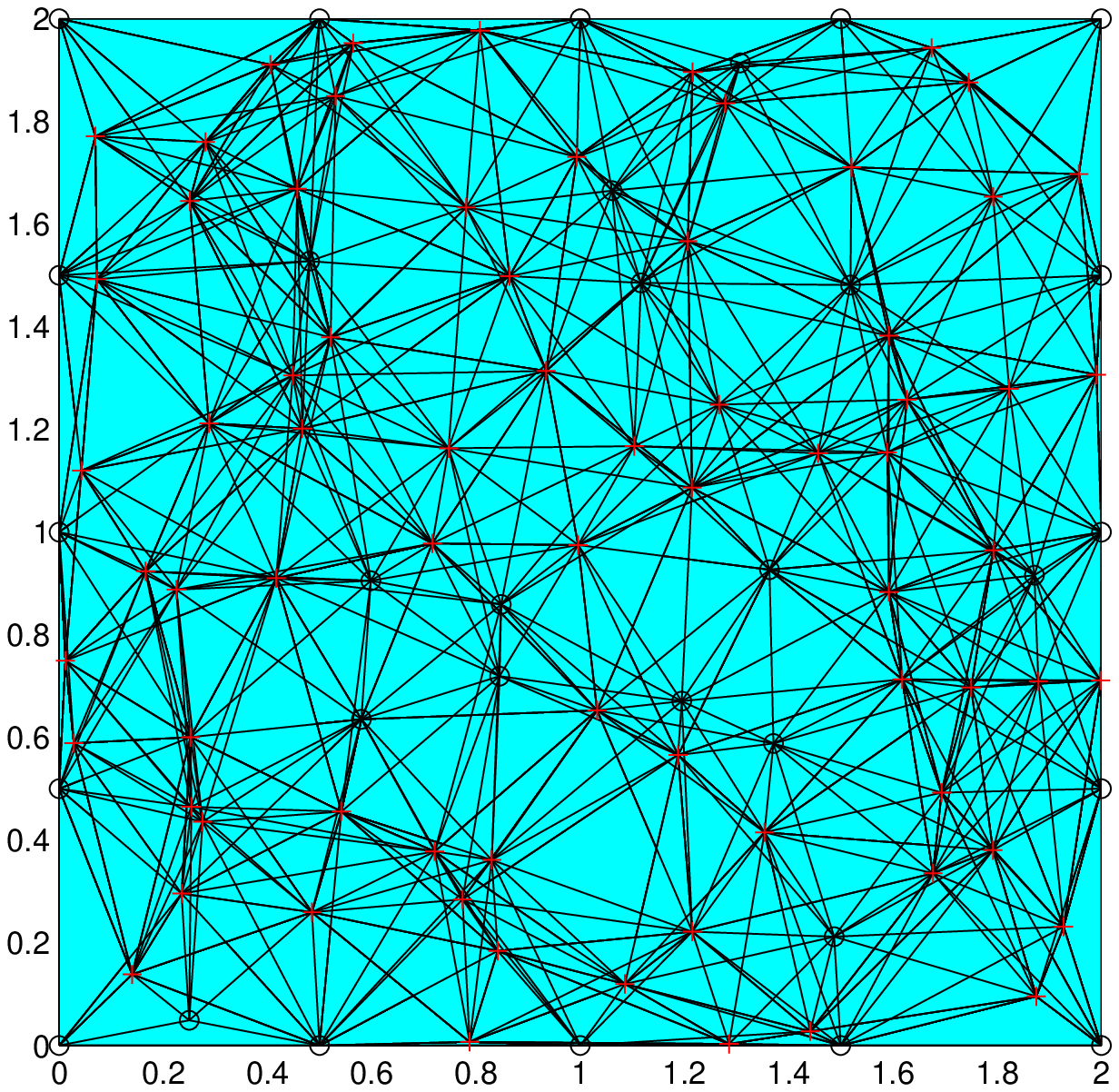}}
\caption{The repaired network.}
    \label{fig_determinantal2}
  \end{figure}

We can see the first step of the disaster recovery algorithm
illustrated in Fig. \ref{fig_determinantal2} for the cellular
network of Fig. \ref{fig_damaged}. Existing nodes are black
circles while added nodes are red plusses.  

\begin{figure}[h]
      \scalebox{0.29}{\includegraphics{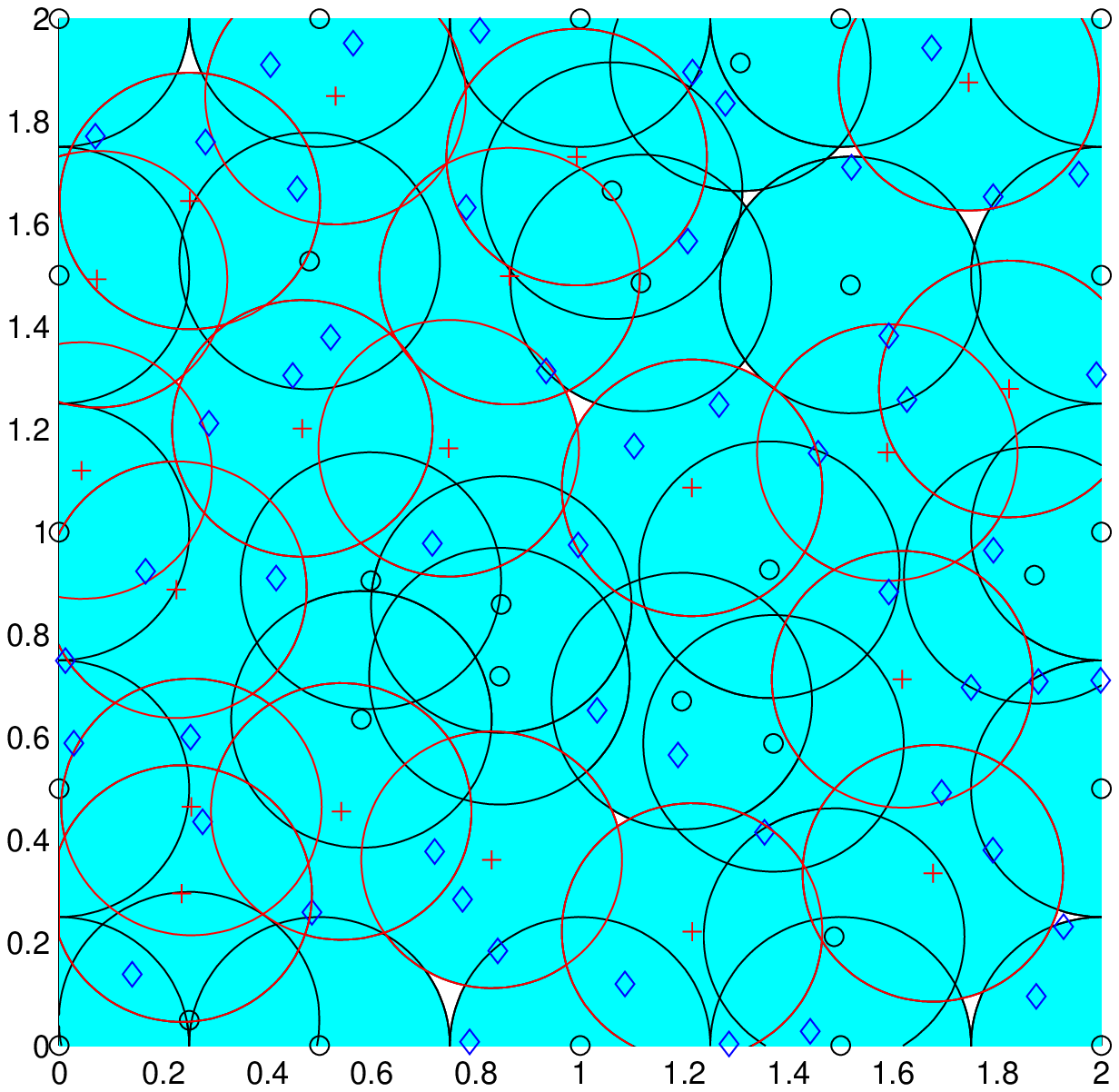}} 
\hfill
      \scalebox{0.29}{\includegraphics{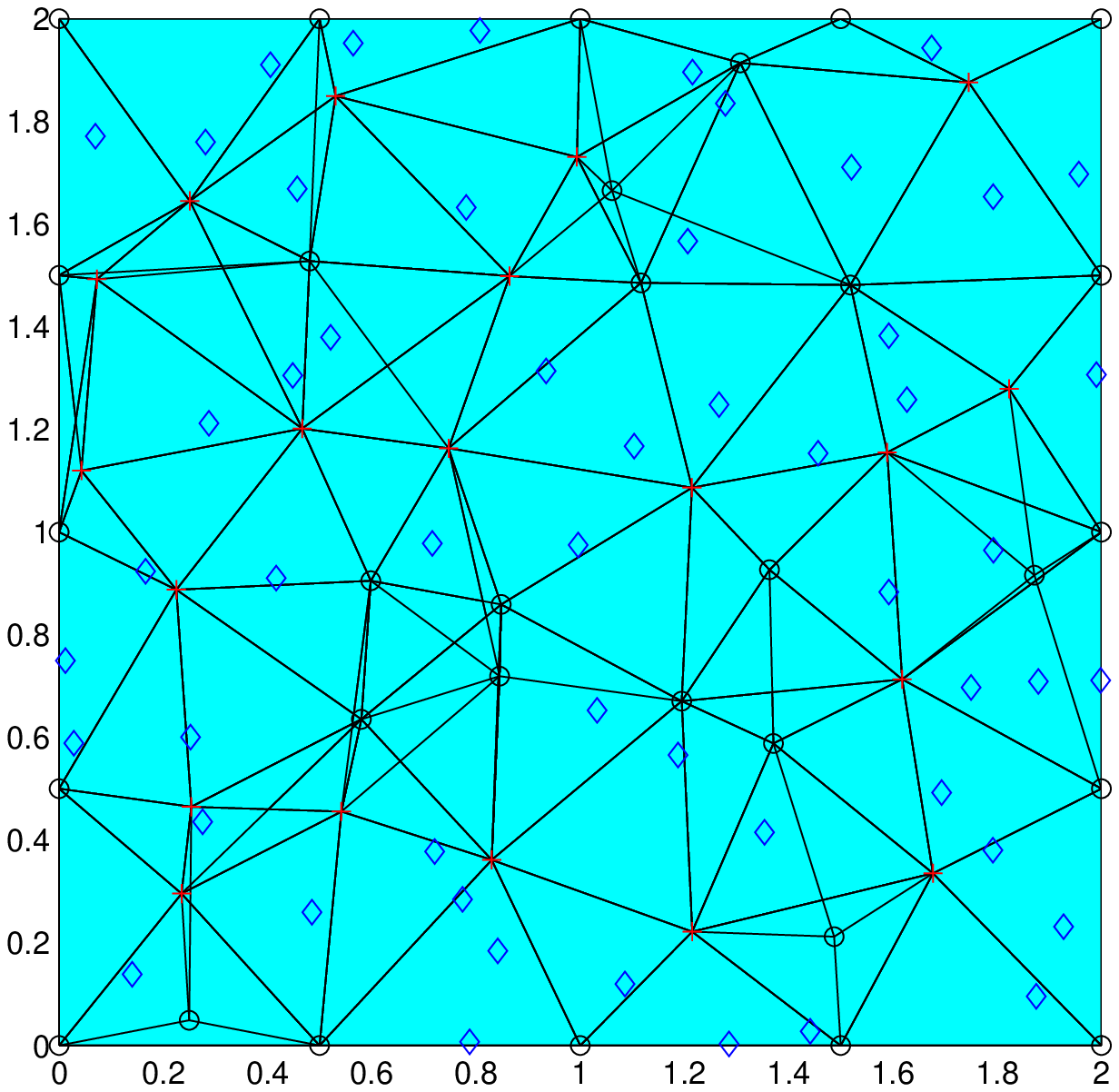}}
    \caption{Final configuration.}
    \label{fig_reduction}
  \end{figure}

Finally, the next step of our approach is to run the reduction
algorithm presented in Section \ref{sec_prel} which maintains the
topology of the repaired cellular  network. At this step, we remove
some of the new nodes we just virtually added in order to achieve an
optimal result with a minimum number of really added nodes.
We can see in Fig. \ref{fig_reduction} an execution of the reduction
algorithm on the intermediate configuration of Figure
\ref{fig_determinantal2} which constitutes of the second and final
step of the disaster recovery algorithm. Removed nodes are represented by blue
diamonds. 

We give in Algorithm \ref{alg_dra} the outline of the algorithm. The
algorithm requires the set of $N_i$ initial nodes $\omega_i$, the fixed
coverage radius $r$, as well as the list of boundary nodes
$L_b$. 
\begin{algorithm}[h]
  \caption{Disaster recovery algorithm}
  \label{alg_dra}
  \begin{algorithmic}[H]
 \REQUIRE Set $\omega_i$ of $N_i$ vertices, for each vertex $v$ its 
neighbor list $L_n(v)$. List of $L_b$ boundary vertices.
    \REQUIRE Set $\omega_i$ of $N_i$ vertices, their coverage radius $r$, list of
    boundary vertices $L_b$
\STATE Computation of the \u{C}ech complex
$X=\mathcal{C}_{r}(\omega)$\ or the Vietoris-Rips complex $X=\mathcal{R}_{2r}(\omega)$\;
\STATE  $N_a=\lceil \frac{a^2}{\pi r^2} \rceil - N_i$\;
\STATE Addition of $N_a$ vertices to $X$ following a determinantal
point process\;
\STATE Computation of $\beta_0(X)$ and $\beta_1(X)$\; 
\STATE $u=1$\;
\WHILE {$\beta_0\neq 1$ \OR $\beta_1 \neq 0$}
\STATE $N_a=N_a+u$\;
\STATE $u=2*u$\;
\STATE Addition of $N_a$ vertices to $X$ following a determinantal
point process\;
\STATE Computation of $\beta_0(X)$ and $\beta_1(X)$\; 
\ENDWHILE
\STATE Reduction algorithm on $X$\;
\RETURN List $L_a$ of kept added vertices.
  \end{algorithmic}
\end{algorithm}

\subsection{Simulation and complexity}
The set of nodes is no more simulated by a Poisson point process,
since we want to simulate a damaged cellular network at different
stages of damage. Every
configuration is still simulated on a square of side $a=2$, and nodes have
a common coverage radius of $r=0.5$ for the construction of the
Vietoris-Rips complex, of which complexity is unchanged $O(N^{N\left(
    \frac{r}{a}^2\right)})$.

The complexity of the drawing of nodes following a determinantal point
process is upper-bounded by the one of the reduction algorithm, see
\cite{ian} for details about the simulation of Ginibre determinantal
point processes. So the complexity of the disaster recovery algorithm
is in $O((1+(\frac{r}{a})^2)^{N})$. It is important to note that the
number of nodes $N$ is the sum of the number of initial nodes $N_i$
and added nodes $N_a$.

\subsection{Performance comparison}
We now compare the performance results of our disaster recovery
algorithm to the best known greedy coverage recovery algorithm: the greedy
algorithm for the set cover problem. As the reduction algorithm, our
homology-based disaster recovery algorithm is of greedy type, so we
compare it to a graph based greedy algorithm. The only way to obtain
coverage with graph representation is to lay nodes following a given
pattern, that is what the greedy algorithm for the set cover problem
does, with a square pattern. It lays a grid of potential new
nodes, and adds first the node the furthest from every other existing
node, and so on until the furthest node is in the coverage of an
existing or added node. 

We first compare the mean number of added nodes following four different
scenarios for the initial state. Indeed, the performance of a recovery
algorithm can be really affected depending on how damaged is the
cellular network to repair. So the different scenarios are defined by
the mean percentage of covered area before running a recovery
algorithm. In Table \ref{table_number}, we present the mean final
number of added nodes for the set cover algorithm and our homology
algorithm over $10^4$ configurations for each scenario. 

\begin{table}[h]
\centering
\begin{tabular}{|ccccc|}
\hline
\% of area initially  covered& 20\% & 40\%& 60\%& 80\%\\ \hline
Set Cover algorithm&$3.65$&$3.36$&$2.82$&$1.84$\\ \hline
Homology algorithm&$4.47$&$3.85$&$2.98$&$1.77$\\ \hline
\end{tabular}
\caption{Mean final number of added nodes $\E{N_f}$}
\label{table_number}
 \end{table}

The numbers of nodes added in the final state both with our recovery
algorithm and the set cover algorithm are roughly the same. Nonetheless,
we can see that our algorithm performs a little bit worse than the set cover
algorithm in the less covered area scenarios. Indeed the set cover
algorithm takes advantage of its grid layout and perfect spacing
between added nodes. However, our homology algorithm gives better
result in the more covered scenarios, thanks to the inherent ability
of determinantal point processes to locate the coverage holes.

In order to show the advantages of our disaster recovery algorithm we choose to
evaluate the compared robustness of the two algorithms when the added nodes positions
are slightly moved, i.e. when the practical positioning does not strictly
follow the theoretical positioning. In order to do this, we apply a
Gaussian perturbation to each the added nodes position. The
covariance matrix of the perturbation is given by $\Sigma=\sigma^2 \mathrm{Id}$ with
$\sigma^2=0.01$, which means that the standard deviation for each
node is of $\sigma=0.1$. Other simulations parameters are unchanged, results
in Table \ref{table_robust} are given in mean over $10^4$
simulations. First, we compute the average number of holes $\E{\beta_1}$
created by the Gaussian perturbation. Then, we count the percentage of simulations
in which the number of holes is still zero, $\P(\beta_1=0)$,  after the Gaussian
perturbation on the new nodes positions.

\begin{table}[h]
\centering
\begin{tabular}{|ccccc|}
\hline
\% of area initially covered&20\% & 40\%& 60\%& 80\%\\ \hline \hline
\multicolumn{5}{|c|}{$\E{\beta_1}$} \\ \hline
Set Cover algorithm&$0.68$&$0.67$&$0.48$&$0.35$\\ \hline
Homology algorithm&$0.58$&$0.52$&$0.36$&$0.26$\\ \hline
\hline
\multicolumn{5}{|c|}{$\P(\beta_1=0)$} \\ \hline
Set Cover algorithm&$40.7\%$&$45.2\%$&$58.8\%$&$68.9\%$\\ \hline
Homology algorithm&$54.0\%$&$58.0\%$&$68.8\%$&$76.1\%$\\ \hline
\end{tabular}
\caption{$\E{\beta_1}$ and $\P(\beta_1=0)$ after the Gaussian perturbation}
\label{table_robust}
 \end{table}

We can see that the perturbation on the number of holes decreases with
the percentage of area initially covered, since the initial nodes
are not perturbed. Our homology algorithm clearly performs better.
Even in the least covered scenarios, there are less than half of the
simulations that create coverage holes, which is not the case for the
set cover algorithm. The set cover algorithm also always create more coverage
holes in mean than our homology algorithm for the same nodes
positions perturbation.
Therefore our algorithm seems more fitted to the disaster
recovery case when a recovery network is deployed in emergency
both indoor, via Femtocells,  and outdoor, via a trailer fleet, where
exact GPS locations are not always available, and exact theoretical
positioning is not always followed. 

\section{Conclusion}
\label{sec_ccl}
In this paper we propose three simplicial homology-based algorithms
that meet three problems of future cellular networks equipped with SON
technology. First, we propose a frequency auto-planning algorithm that
uses simplicial homology to provide a homogeneous usage of
frequencies. Then, we present an energy conservation algorithm that
adapt the stay-on nodes of a cellular network to both coverage,
computed via simplicial homology, and the capacity defined by user demand. Finally, we
introduce a disaster recovery algorithm in which we use simplicial
homology to compute coverage, and the repulsive Ginibre determinantal
point process to propose an efficient fixing to a damaged network.

We investigate the performance of these three homology-based
algorithm. We compute their complexity that is highly dependent on the
simplicial complex representation which is necessary to the topology
computation. We compare our three homology-based greedy
algorithms to three graph-based greedy algorithms to exhibit the
benefits of using simplicial homology. We show the
advantages of our algorithms, which are not mathematical optimality
but application oriented. Indeed, some tradeoffs in complexity and
performance are necessary in order to obtain results more suited to
the applications encountered in future cellular networks. Despite the
fact that the optimal number of frequencies is not always reached, our 
 frequency auto-planning algorithm provides an even and better
 coverage for every planned frequency than traditionnal methods. Our energy
conservation algorithm is the first  to take into account both user traffic and
coverage. And even if a square grid method is mathematically optimal,
our disaster recovery algorithm offers a repaired network 
robust to slight modifications of its nodes positions and is able to
target coverage holes.

As of future work, it is first possible to consider a more complex
model for the cellular network where some nodes location are
known. Indeed, with the arrival of HetNet, every node locations is not
always node, however the operator has still full knowledge of the eNBs
locations. Then it would be possible to mix a homological approach to
a geometrical approach for the coverage computation, and some gains in
complexity could be achieved. Secondly, a next step is to derive more
advanced algorithms from our homology-based approach. Gains in both
complexity and performance are likely, and comparison with more
elaborate graph-based algorithm would be possible.
\bibliographystyle{IEEEtran}
\bibliography{article}

\begin{thebibliography}{10}
\providecommand{\url}[1]{#1}
\csname url@samestyle\endcsname
\providecommand{\newblock}{\relax}
\providecommand{\bibinfo}[2]{#2}
\providecommand{\BIBentrySTDinterwordspacing}{\spaceskip=0pt\relax}
\providecommand{\BIBentryALTinterwordstretchfactor}{4}
\providecommand{\BIBentryALTinterwordspacing}{\spaceskip=\fontdimen2\font plus
\BIBentryALTinterwordstretchfactor\fontdimen3\font minus
  \fontdimen4\font\relax}
\providecommand{\BIBforeignlanguage}[2]{{%
\expandafter\ifx\csname l@#1\endcsname\relax
\typeout{** WARNING: IEEEtran.bst: No hyphenation pattern has been}%
\typeout{** loaded for the language `#1'. Using the pattern for}%
\typeout{** the default language instead.}%
\else
\language=\csname l@#1\endcsname
\fi
#2}}
\providecommand{\BIBdecl}{\relax}
\BIBdecl

\bibitem{normelte}
3GPP, ``T{R} 36.913: Requirements for further advancements for evolved
  universal terrestrial radio access ({E-UTRA}),'' \emph{Tech. Rep.}, Mar.
  2009.

\bibitem{bookson}
S.~Hmlinen, H.~Sanneck, and C.~Sartori, \emph{{LTE} Self-Organising Networks
  ({SON}): Network Management Automation for Operational Efficiency},
  1st~ed.\hskip 1em plus 0.5em minus 0.4em\relax Wiley Publishing, 2012.

\bibitem{cognitive}
S.~Haykin, ``Cognitive radio: brain-empowered wireless communications,''
  \emph{Selected Areas in Communications, IEEE Journal on}, vol.~23, no.~2, pp.
  201--220, 2005.

\bibitem{ofdmafemto}
D.~Lopez-Perez, A.~Valcarce, G.~de~la Roche, and J.~Zhang, ``Ofdma femtocells:
  A roadmap on interference avoidance,'' \emph{Communications Magazine, IEEE},
  vol.~47, no.~9, pp. 41--48, September 2009.

\bibitem{infocom}
A.~Vergne, L.~Decreusefond, and P.~Martins, ``Reduction algorithm for
  simplicial complexes,'' in \emph{INFOCOM, 2013 Proceedings IEEE}, 2013, pp.
  475--479.

\bibitem{pmr}
A.~Vergne, I.~Flint, L.~Decreusefond, and P.~Martins,
  ``\BIBforeignlanguage{Anglais}{{Disaster Recovery in Wireless Networks: A
  Homology-Based Algorithm}},'' in \emph{\BIBforeignlanguage{Anglais}{ICT
  2014}}, May 2014.

\bibitem{jantti}
J.~Li and R.~Jantti, ``On the study of self-configuration neighbour cell list
  for mobile {W}i{MAX},'' \emph{Next Generation Mobile Applications, Services
  and Technologies, International Conference on}, vol.~0, pp. 199--204, 2007.

\bibitem{kim}
D.~Kim, B.~Shin, D.~Hong, and J.~Lim, ``Self-configuration of neighbor cell
  list utilizing {E-UTRAN} node{B} scanning in {LTE} systems,'' in
  \emph{Consumer Communications and Networking Conference (CCNC), 2010 7th
  IEEE}, 2010, pp. 1--5.

\bibitem{irani}
\BIBentryALTinterwordspacing
S.~Irani, ``\BIBforeignlanguage{English}{Coloring inductive graphs on-line},''
  \emph{\BIBforeignlanguage{English}{Algorithmica}}, vol.~11, no.~1, pp.
  53--72, 1994. [Online]. Available: \url{http://dx.doi.org/10.1007/BF01294263}
\BIBentrySTDinterwordspacing

\bibitem{brian}
A.~Nasif and B.~Mark, ``Opportunistic spectrum sharing with multiple cochannel
  primary transmitters,'' \emph{Wireless Communications, IEEE Transactions on},
  vol.~8, no.~11, pp. 5702--5710, 2009.

\bibitem{das}
S.~Das, S.~Sen, and R.~Jayaram, ``A structured channel borrowing scheme for
  dynamic load balancing in cellular networks,'' in \emph{Distributed Computing
  Systems, 1997., Proceedings of the 17th International Conference on}, 1997,
  pp. 116--123.

\bibitem{agha}
T.~Al-Meshhadany and K.~Al-Agha, ``{VCB} by means of soft2hard handover in
  {WCDMA},'' in \emph{Mobile and Wireless Communications Network, 2002. 4th
  International Workshop on}, 2002, pp. 487--491.

\bibitem{fujii}
T.~Fujii and S.~Nishioka, ``Selective handover for traffic balance in mobile
  radio communications,'' in \emph{Communications, 1992. ICC '92, Conference
  record, SUPERCOMM/ICC '92, Discovering a New World of Communications., IEEE
  International Conference on}, 1992, pp. 1840--1846 vol.4.

\bibitem{du}
L.~Du, J.~Bigham, and L.~Cuthbert, ``An intelligent geographic load balance
  scheme for mobile cellular networks,'' in \emph{Computer Communications and
  Networks, 2002. Proceedings. Eleventh International Conference on}, 2002, pp.
  348--353.

\bibitem{rittenhouse}
S.~Das, H.~Viswanathan, and G.~Rittenhouse, ``Dynamic load balancing through
  coordinated scheduling in packet data systems,'' in \emph{INFOCOM 2003.
  Twenty-Second Annual Joint Conference of the IEEE Computer and
  Communications. IEEE Societies}, vol.~1, 2003, pp. 786--796 vol.1.

\bibitem{peng}
\BIBentryALTinterwordspacing
P.~Jiang, J.~Bigham, and J.~Wu, ``Self-organizing relay stations in relay based
  cellular networks,'' \emph{Comput. Commun.}, vol.~31, no.~13, pp. 2937--2945,
  Aug. 2008. [Online]. Available:
  \url{http://dx.doi.org/10.1016/j.comcom.2008.02.024}
\BIBentrySTDinterwordspacing

\bibitem{claussen}
I.~Ashraf, L.~T.~W. Ho, and H.~Claussen, ``Improving energy efficiency of
  femtocell base stations via user activity detection,'' in \emph{Wireless
  Communications and Networking Conference (WCNC), 2010 IEEE}, 2010, pp. 1--5.

\bibitem{nash}
\BIBentryALTinterwordspacing
E.~Campos-Na{\~n}ez, A.~Garcia, and C.~Li, ``A game-theoretic approach to
  efficient power management in sensor networks,'' \emph{Oper. Res.}, vol.~56,
  no.~3, pp. 552--561, 2008. [Online]. Available:
  \url{http://dx.doi.org/10.1287/opre.1070.0435}
\BIBentrySTDinterwordspacing

\bibitem{kaschub}
\BIBentryALTinterwordspacing
C.~Mueller, M.~Kaschub, C.~Blankenhorn, and S.~Wanke, ``A cell outage detection
  algorithm using neighbor cell list reports,'' in \emph{Self-Organizing
  Systems}, ser. Lecture Notes in Computer Science, K.~Hummel and J.~Sterbenz,
  Eds.\hskip 1em plus 0.5em minus 0.4em\relax Springer Berlin Heidelberg, 2008,
  vol. 5343, pp. 218--229. [Online]. Available:
  \url{http://dx.doi.org/10.1007/978-3-540-92157-8_19}
\BIBentrySTDinterwordspacing

\bibitem{amirijoo}
M.~Amirijoo, L.~Jorguseski, T.~Kurner, R.~Litjens, M.~Neuland, L.~Schmelz, and
  U.~Turke, ``Cell outage management in {LTE} networks,'' in \emph{Wireless
  Communication Systems, 2009. ISWCS 2009. 6th International Symposium on},
  2009, pp. 600--604.

\bibitem{selfhealing}
3GPP, ``Telecommunications management; self-healing {OAM}; concepts and
  requirements,'' \emph{Tech. Rep.}, vol. 3GPP TS 32.541 v1.6.1, 2010.

\bibitem{att}
K.~Morrison, ``Rapidly recovering from the catastrophic loss of a major
  telecommunications office,'' \emph{Communications Magazine, IEEE}, vol.~49,
  no.~1, pp. 28--35, 2011.

\bibitem{greedy}
\BIBentryALTinterwordspacing
V.~Chvatal, ``{A Greedy Heuristic for the Set-Covering Problem},''
  \emph{Mathematics of Operations Research}, vol.~4, no.~3, pp. 233--235, 1979.
  [Online]. Available: \url{http://dx.doi.org/10.2307/3689577}
\BIBentrySTDinterwordspacing

\bibitem{epsilon-nets}
\BIBentryALTinterwordspacing
D.~Haussler and E.~Welzl, ``Epsilon-nets and simplex range queries,'' in
  \emph{Proceedings of the second annual symposium on Computational geometry},
  ser. SCG '86.\hskip 1em plus 0.5em minus 0.4em\relax New York, NY, USA: ACM,
  1986, pp. 61--71. [Online]. Available:
  \url{http://doi.acm.org/10.1145/10515.10522}
\BIBentrySTDinterwordspacing

\bibitem{fps2}
Q.~Fang, J.~Gao, L.~Guibas, V.~de~Silva, and L.~Zhang, ``{GLIDER}: Gradient
  landmark-based distributed routing for sensor networks,'' in \emph{Proc. IEEE
  Conference on Computer Communications (INFOCOM)}, 2005.

\bibitem{fps1}
A.Nguyen, N.Milosavljevi\'c, Q.Fang, J.Gao, and L.J.Guibas, ``Landmark
  selection and greedy landmark-descent routing for sensor networks,'' in
  \emph{Proceedings of IEEE INFOCOM 2007}, 2007.

\bibitem{surveyson}
O.~Aliu, A.~Imran, M.~Imran, and B.~Evans, ``A survey of self organisation in
  future cellular networks,'' \emph{Communications Surveys Tutorials, IEEE},
  vol.~15, no.~1, pp. 336--361, 2013.

\bibitem{hatcher}
A.~Hatcher, \emph{Algebraic Topology}.\hskip 1em plus 0.5em minus 0.4em\relax
  Cambridge University Press, 2002.

\bibitem{Feng2}
F.~{Yan}, P.~{Martins}, and L.~{Decreusefond}, ``Accuracy of homology based
  approaches for coverage hole detection in wireless sensor networks,'' in
  \emph{ICC 2012}, Jun. 2012.

\bibitem{hauteur}
\BIBentryALTinterwordspacing
L.~Decreusefond, P.~Martins, and A.~Vergne,
  ``\BIBforeignlanguage{Anglais}{{Reduction algorithm for random abstract
  simplicial complexes}},'' Mar. 2014, hal-00864303. [Online]. Available:
  \url{http://hal.archives-ouvertes.fr/hal-00864303}
\BIBentrySTDinterwordspacing

\bibitem{ian}
\BIBentryALTinterwordspacing
L.~Decreusefond, I.~Flint, and A.~Vergne,
  ``\BIBforeignlanguage{Anglais}{{Efficient simulation of the Ginibre
  process}},'' Oct. 2013, hal-00869259. [Online]. Available:
  \url{http://hal.archives-ouvertes.fr/hal-00869259}
\BIBentrySTDinterwordspacing

\end{thebibliography}
\vspace{-10mm}
\begin{IEEEbiography}[{\includegraphics[width=1in,height=1.25in,clip,keepaspectratio]{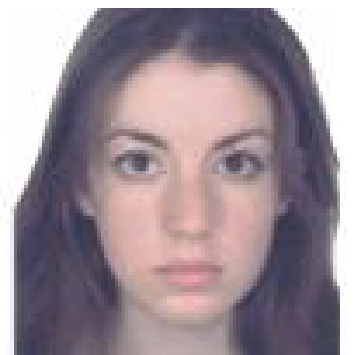}}]{Ana\"is Vergne}
received the Dipl.Ing. degree in telecommunications from
Telecom ParisTech, Paris, France in 2010. She obtained the
Ph.D. degree in networking and computer sciences in 2013 from Telecom
ParisTech, Paris, France. She is currently a post-doctoral fellow in
the Geometrica team at Inria Saclay - Ile de France, Palaiseau, France.
Her research interests include
stochastic geometry applications to wireless networks, more
particularly algebraic topology applied to wireless sensor networks.
\end{IEEEbiography}
\vspace{-10 mm}
\begin{IEEEbiography}[{\includegraphics[width=1in,height=1.25in,clip,keepaspectratio]{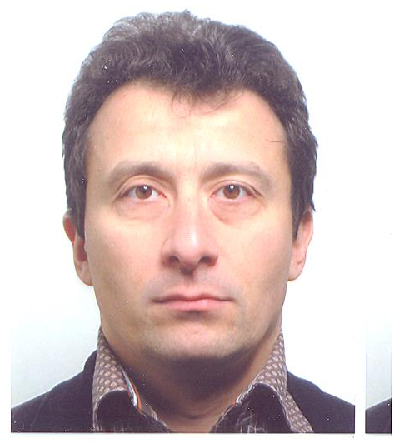}}]{Laurent Decreusefond}
is a former student of Ecole Normale Supérieure
de Cachan. He obtained his Ph.D. degree in Mathematics in 1994 from
Telecom ParisTech and his Habilitation in 2001. He is currently a
Professor in the Network and Computer Science Department, at Telecom
ParisTech. His main fields of interest are the Malliavin calculus, the
stochastic analysis of long range dependent processes, random geometry
and topology and their applications. With P. Moyal, he co-authored a
book about the stochastic modeling of telecommunication.
\end{IEEEbiography}
\vspace{-10 mm}
\begin{IEEEbiography}[{\includegraphics[width=1in,height=1.25in,clip,keepaspectratio]{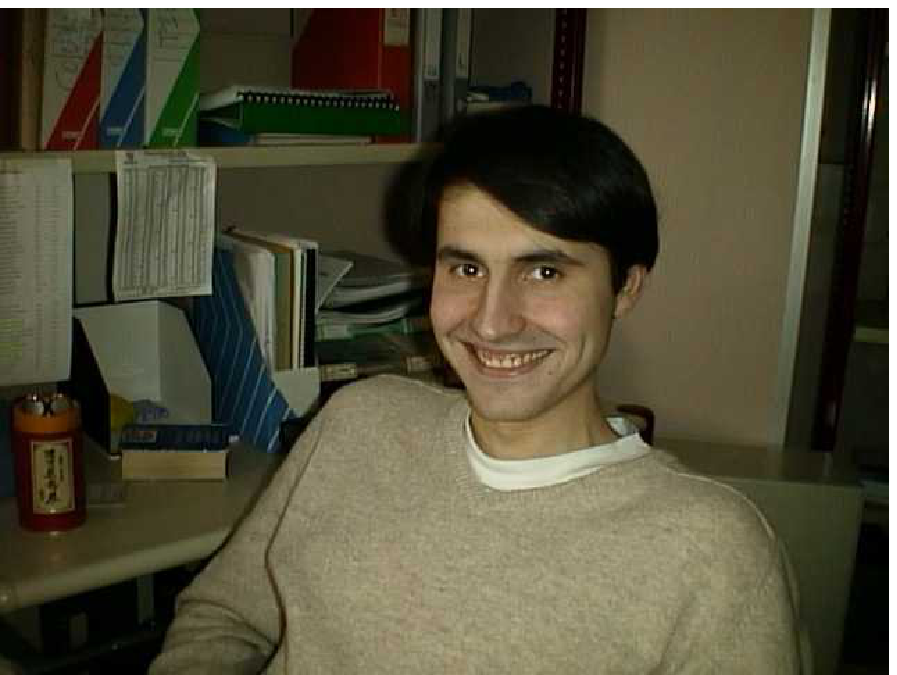}}]{Philippe Martins}
received a M.S. degree in signal
processing and another M.S. degree in networking
and computer science from Orsay University and
ESIGETEL France, in 1996. He received the Ph.D.
degree in electrical engineering from
Telecom ParisTech, Paris, France, in 2000.
He is currently a Professor in the Network and
Computer Science Department, at Telecom Paris-
Tech. His main research interests lie in performance
evaluation in wireless networks (RRM, scheduling,
handover algorithms, radio metrology). His current
investigations address mainly three issues: a) the design of distributed
sensing
algorithms for cognitive radio b) distributed coverage holes detection in
wireless sensor networks c) the definition of analytical models for the
planning
and the dimensioning of cellular systems. He has published several papers on
different international journals and conferences. He is also an IEEE senior
member and he is co-author of several books on 3G and 4G systems.
\end{IEEEbiography}

\end{document}